\documentclass[aps,twocolumn,showpacs,superscriptaddress,pre,floatfix]{revtex4-2}
\usepackage{amsmath}
\usepackage{amssymb}
\usepackage{dcolumn}%

\usepackage{tikz}
\usepackage{mathtools}
\usepackage{enumitem}
\usepackage{hyperref}
\hypersetup{
    colorlinks,
    linkcolor={blue!80!black},
    citecolor={blue!80!black},
    urlcolor={blue!80!black}
}
\usepackage{multirow}
\usepackage{amsthm}
\usepackage{placeins}

\def\rv{J_2}
\def\nn{NN}
\def\nnn{NNN}
\def\resultZeroTEntropyMinusQuarter{0.230\,960\,93(14)}
\begin{document}

\title{Meandering stripes in the frustrated $J_1$-$J_2$ Ising model on the honeycomb lattice}
\date{\today}

\author{Denis Gessert}
\affiliation{Centre for Fluid and Complex Systems, Coventry University, Coventry, CV1 5FB, United Kingdom}
\affiliation{Institut f\"{u}r Theoretische Physik, Universit\"at Leipzig, IPF 231101, 04081 Leipzig, 
  Germany}
\affiliation{Institut f\"ur Physik, Technische Universit\"at Chemnitz, 09107 Chemnitz, Germany}

\author{Martin Weigel}
\affiliation{Institut f\"ur Physik, Technische Universit\"at Chemnitz, 09107 Chemnitz, Germany}

\author{Wolfhard Janke}
\affiliation{Institut f\"{u}r Theoretische Physik, Universit\"at Leipzig, IPF 231101, 04081 Leipzig, 
  Germany}

\begin{abstract}
	We study the frustrated $J_1$-$J_2$ Ising model on the honeycomb lattice with ferromagnetic nearest-neighbor couplings fixed at $J_1=1$ and strong antiferromagnetic next-nearest-neighbor interactions, i.e., $J_2 \leq -1/4$. Little is known for this range of $J_2$, whereas for less negative values of $J_2$ the system orders ferromagnetically at low temperatures and appears to remain in the Ising universality class. In previous work it was shown that the model has a largely degenerate ground state, and it was conjectured that there is some kind of phase transition. We introduce a complex-valued nematic order parameter, which can differentiate between the high-temperature paramagnetic phase and the observed partially-disordered stripe phase at lower temperatures. Configurations in this phase consist of stripes of spins parallel with respect to one lattice direction, which collectively meander along the remaining two, producing partially disordered ground states. The sharp peaks in the specific heat observed in earlier work only appear when using periodic boundary conditions and are absent for free boundaries. Additionally, we reveal a striking dependence of the behavior on the aspect ratio of the considered samples. Ultimately, even a careful finite-size scaling analysis for $J_2 = -0.5$ and $J_2 = -1$ is unable to clearly discern between a crossover without any singularities and some form of continuous transition, including the possibility of an infinite-order transition of the Berezinskii-Kosterlitz-Thouless (BKT) type. For $J_2=-1/4$ we find that the system remains disordered at all temperatures and that it exhibits a finite ground-state entropy per site, for which our simulations provide the accurate asymptotic estimate $S(T=0)/N = \resultZeroTEntropyMinusQuarter$ in the thermodynamic limit $N\rightarrow\infty$.
\end{abstract}

\maketitle

\section{Introduction}

\begin{figure*}[tb!]
\centering
  \includegraphics[scale=1.1]{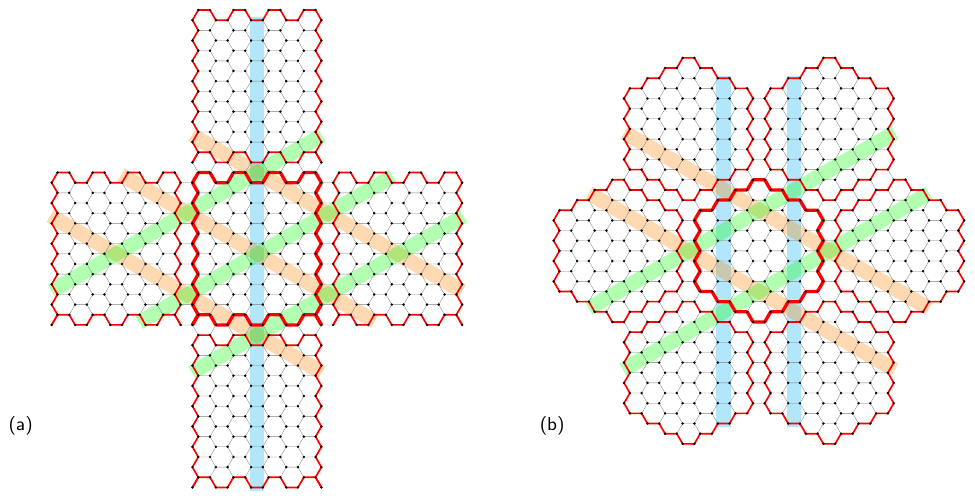}
\caption{Honeycomb lattice with (a) rectangular and (b) hexagonal periodic boundary conditions. Black circles represent spin sites, and dashed lines correspond to nearest-neighbor interactions. Solid red lines show the system's boundaries. For both (a) and (b), $L=8$. When one of the three lattice axes is pinned, a wrapping line of pairs of spins in the direction of that axis can be flipped without changing the energy. The cyan, green, and orange regions show a single exemplary line for the axes 1, 2, and 3, respectively.}\label{fig:honeycombLattice} 
\end{figure*}

Frustrated spin systems exhibit a strikingly wide spectrum of behaviors~\cite{Diep2012}, such as reentrance~\cite{Azaria1987}, degenerate ground states~\cite{Wannier1950,*Wannier1973Errata, Espriu2004}, and several consecutive phase transitions for positive temperatures~\cite{vaks1966ising}. One of the simplest such models is the $J_1$-$J_2$ Ising model on the square lattice with competing ferromagnetic nearest-neighbor~($J_1>0$) and antiferromagnetic next-nearest-neighbor interactions~($J_2<0$), which has been extensively studied over the past decades~\cite{Binder1980,Bloete1987, Malakis2006, Kalz2008, Kalz2011, Kalz2012, Jin2012}. By now, it may be considered relatively well understood, although the debate on some questions remains ongoing~\cite{Hu2021,Li2021a,Yoshiyama2023,Chatelain2025}.

More recently, the same model on the honeycomb lattice has started to attract significant interest~\cite{Bobak2016,Zukovic2020,Zukovic2021,Schmidt2021,Acevedo2021, Corte2021,Zukovic2022,Dias2023,Gessert2024,Gessert2025,Azhari2025,Batista2026}, and this system is the focus of the present work. We previously reported results for this model for $J_2 / J_1 > -1/4$, where it orders ferromagnetically at low temperatures, and where it remains in the Ising universality class (although we did not present simulation data for $-1/4 < J_2/|J_1| < -0.23$)~\cite{Gessert2024,Gessert2025}. Much less is known for the regime of strong antiferromagnetic next-nearest-neighbor couplings, i.e., $J_2/J_1\le -1/4$. The ground state in this regime is infinitely degenerate~\cite{Kudo1976,Katsura1986}. Using effective field theory, no long-range order was detected~\cite{Bobak2016}, whereas Monte Carlo simulations~\cite{Zukovic2020,Acevedo2021,Zukovic2022,Azhari2025} indicate some kind of transition, show very slow relaxation reminiscent of spin glasses~\cite{Zukovic2020}, and for some coupling ratios $J_1/J_2$ multiple peaks in the specific heat were observed, some of which appear to show behavior reminiscent of a first-order transition~\cite{Zukovic2022}.
We also note a study of this model employing unsupervised machine learning techniques~\cite{Acevedo2021}, which finds a signal in the reconstruction error indicative of a phase transition in this regime (similar to findings in the ferromagnetic region for which the existence of a phase transition is well established).
A recent Wang-Landau simulation study~\cite{Azhari2025}, focusing on the specific-heat peak with the highest temperature, indicates that the peak might be associated with a thermodynamic transition. The authors find it to be of first-order for $J_2 / J_1 \geq -0.3$ and of second order for $J_2 / J_1 \leq -0.5$ --- with a tricritical point $J_2^{*} \in (-0.5 J_1,-0.3 J_1)$ separating the two regimes. There is hence a wealth of potentially conflicting results for the $J_1$-$J_2$ honeycomb Ising model in this regime.

Besides the difficulty of equilibrating the system in simulations, the lack of an order parameter describing the observed transition further hinders the understanding of the temperature behavior and phase diagram.  A periodic system with linear extent $L$ has $2^{O(L)}$ degenerate ground states~\cite{Azhari2025} (unless $J_2 / J_1 = -1/4$, where the degeneracy is extensive).  We identify a symmetry that is broken within this set of ground states, and we define a nematic order parameter $\eta$ that is sensitive to such symmetry breaking.  As we will see, this symmetry is affected by the type of boundary conditions used, which is why we study four different types of boundaries.  We note that even for much simpler models such as the Ising antiferromagnet (AFM) on a triangular lattice the relevant behavior depends on the chosen boundary conditions~\cite{Millane2004,Millane2006,Kim2015}, which further motivated us to extend the study to cover several plausible boundary conditions.  Doing so allows us to differentiate between boundary-condition-dependent and -independent features. Additionally, we observe a marked dependence of results on the aspect ratio of the samples considered, which is another effect that is unfamiliar from systems without frustration.

The rest of the paper is organized as follows. The model, its ground-state configurations, the topological defects arising from thermal excitations, and the measured observables (including the proposed order parameter $\eta$) are introduced in Sec.~\ref{sec:modelObservables}. In Sec.~\ref{sec:algo} we present details regarding the simulation setup and algorithm. Our results are presented in Sec.~\ref{sec:results} and discussed in Sec.~\ref{sec:discussion}. Finally, we summarize our findings and provide some outlook in Sec.~\ref{sec:conclusion}. Several more technical aspects are discussed in the Appendices \ref{app:energyMinFBC}--\ref{app:analogousBehaviorInSimpleModels}.

\section{Model and observables}\label{sec:modelObservables} \subsection{Model and boundary conditions}\label{sec:modelAndBC}

The model is described by the following spin Hamiltonian: \begin{equation} \mathcal{H} = -J_1 \sum_{\langle ij \rangle} \sigma_i \sigma_j - J_2 \sum_{[ ik ]} \sigma_i \sigma_k,\label{eq:Hamiltonian} \end{equation} where $\sigma_i$ are Ising spins on the honeycomb lattice, $\sum_{\langle ij \rangle}$ refers to the sum over nearest neighbors (\nn{}) and $\sum_{[ ik ]}$ to the one over next-nearest neighbors (\nnn{}), respectively.  To fix units, we choose $J_1=1$ and $k_\mathrm{B}=1$.  Note that the same model with antiferromagnetic $J_1$ has alternatively been studied instead~\cite{Bobak2016,Zukovic2020,Zukovic2021,Zukovic2022,Dias2023}. As the nearest-neighbor structure is bipartite, positive and negative $J_1$ can be mapped onto each other by a simple transformation. Hence, all statements regarding the phase diagram apply \emph{mutatis mutandis} to the fully antiferromagnetic version of the model as well.

	Since boundary effects play a crucial role in anti-ferromagnetic and frustrated models~\cite{Millane2004,Millane2006,Kim2015}, we consider the model both using periodic and free boundary conditions. As for the shape of the sample, a natural choice appears to be a rectangular system, yielding rectangular periodic boundary conditions (PBC) and rectangular free boundary conditions (FBC), see Fig.~\ref{fig:honeycombLattice}(a). In Fig.~\ref{fig:honeycombLattice}, the actual spin configuration corresponds to the central region enclosed by the thick red lines. Periodic boundary conditions can then be imagined as placing identical copies of the system along each edge of the central system. %
  The system depicted in Fig.~\ref{fig:honeycombLattice}(a) is considered to have a linear extent of $L=8$ since it contains $L^2$ hexagons with $N=(6/3)L^2=2 L^2$ spins at their vertices. (The factor of two can also be understood as reflecting the fact that there are two sites in a unit cell of the honeycomb lattice.)
  Note that in both cases $L$ needs to be even in order to allow for the periodic repetition of the lattice.

	However, the threefold (rotational) symmetry of the lattice is not respected when embedding the system in a rectangular shape: As can be seen from the wrapping lines in Fig.~\ref{fig:honeycombLattice}(a), two out of three principal lattice axes are equivalent (axes 2 and 3 in green and orange), while axis 1 (cyan) differs.  We therefore also study hexagonal periodic boundary conditions as well as hexagonal free boundaries, see Fig.~\ref{fig:honeycombLattice}(b) where we again highlight the three principal directions, which are now found to be equivalent, as expected. In this case, $L$ denotes the length of each of the six zigzag edges [see Fig.~\ref{fig:honeycombLattice}(b) showing an $L=8$ system], that is there are $L/2$ hexagons on each edge. In such a system there are $N=3 L^2/2$ sites in total

	For free boundary conditions on the honeycomb lattice, two types of edges can arise, namely armchair edges and zigzag edges~\footnote{The two terms `armchair' and `zigzag' are widely used in the context of graphene nanoribbons, graphene of course being the most prominent material with a honeycomb structure.}.  For the rectangular geometry in Fig.~\ref{fig:honeycombLattice}(a) the former are present at the bottom and top boundaries, whereas the latter can be found on the left and right edges. For the hexagonal case depicted in Fig.~\ref{fig:honeycombLattice}(b) all boundaries are zigzag edges.  As the spin neighborhoods of surface spins are different on zigzag and armchair edges, surface spin configurations that minimize the energy are also different from each other and from the ones in the bulk; for more details, we refer to the next section as well as to Appendix~\ref{app:energyMinFBC}.

	We also note that different aspect ratios of rectangular samples have been studied in the past, usually arguing that in some representation they appear to be the closest approximation to a square shape. While for an unfrustrated system the influence of such a change on thermodynamic observables usually is minor, this is not the case for the frustrated $J_1$--$J_2$ Ising model and, instead, we find some rather profound effects that can be traced back to different aspect ratios of the considered systems. The effect of varying the aspect ratio is discussed in Sec.~\ref{sec:aspectRatio} below.

	\subsection{Energy minimization and ground states}\label{sec:groundStates}

	Disregarding possible surface effects when using free boundary conditions, the ground-state energy per spin can be determined by local energy minimization. The lowest energy is attained by the ferromagnetic ground state with energy
  \begin{equation}
     e_\mathrm{FM} = - \frac 3 2 (1 + 2 J_2),
  \end{equation}
  for $\rv > -1/4$, or by striped configurations with energy
  \begin{equation}
     e_\mathrm{S} = -\frac 1 2 ( 1 - 2 J_2)
  \end{equation}
  for $\rv<-1/4$~\cite{Katsura1986,Bobak2016}. It is easy to see that for the special point $\rv = -1/4$, the two energies are equal, which gives rise to a highly degenerate ground state, which we will study in Sec.~\ref{sec:minusOneQuarter}. The nature of the stripe states for $\rv < -1/4$, which are the only ones relevant at low temperatures in the phase studied here, will be discussed in the following. %
  These states are characterized by the fact that two out of the three {\nn} interactions and four out of the six {\nnn} interactions are satisfied, leaving the remaining interactions frustrated. This frustration leads to a highly degenerate ground state, as was observed in earlier work~\cite{Houtappel1950,Zukovic2022,Azhari2025}.  In the limit $\rv\rightarrow-\infty$ the system becomes equivalent to two decoupled antiferromagnetic Ising models on triangular lattices (cf.\ Fig.~\ref{fig:boundaryConfigs} as well as Fig.~1 of Ref.~\cite{Gessert2025}). For the triangular Ising AFM its extensive ground-state entropy is known exactly, as well as that there is no phase transition~\cite{Wannier1950,*Wannier1973Errata}. Hence, the same is the case for the present model in the limit $J_2 \to -\infty$.

\begin{figure}[tb!]
  \includegraphics{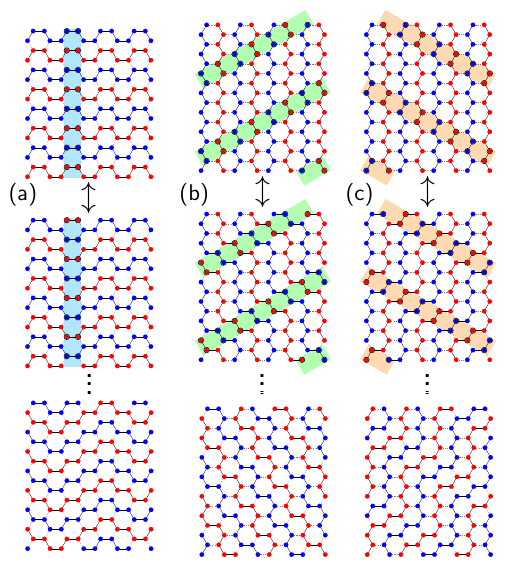}
  \caption{Exemplary wrapping lines of pairs of spins that can be flipped without changing the energy when the respective axis is pinned. Panels (a), (b) and (c) correspond to the axes 1, 2 and 3, respectively. The top row shows possible starting configurations, the middle row indicates moves to generate different ground states, and the bottom row shows random ground states generated in this fashion.\label{fig:groundstateMoves}}
\end{figure}

Since for $-\infty<J_2<-1/4$ the energy is minimized by two of the neighboring spins having the same spin value and one pointing in the opposite direction, the ground states naturally consist of stripes of parallel spins. These stripes cannot intersect without an increase in energy, such that they have to be parallel with respect to one of the principal lattice directions.  After pinning one of the three axes, stripes collectively meander along the two remaining directions, cf.\ Fig.~\ref{fig:groundstateMoves}. Hence, the stripes follow a random-walk (RW) like pattern, giving rise to a large degeneracy at zero temperature, and hence the ground states remain partially disordered. At higher temperatures, stripes are only locally parallel and regions, in which different directions are pinned, are connected by topological defects (see Sec.~\ref{sec:topologicalDefects}).  The existence of parallel stripes effectively breaks the threefold symmetry of the lattice, an effect that can be measured by the nematic order parameter introduced below in Sec.~\ref{sec:observables}.

In the case of \emph{periodic boundary conditions}, a ground state can be described by a single such RW-like stripe, which then is replicated along the pinned direction with alternating spin orientations. The ``initial'' stripe has a length of the order of $L$, and it can turn in two different directions at each hexagon it passes. Thus, there are $2^{O(L)}$ many realizations of such stripes, and the zero-temperature entropy scales linearly in $L$.  As for the actual number of ground states for a given system, this depends non-trivially on the system's embedding, including its aspect ratio.

The precise number of ground states can be determined as follows. Consider a system of size $L$ and \emph{rectangular periodic boundary conditions} such as the one depicted in Fig.~\ref{fig:honeycombLattice}(a).  Since each ground state can be described by a RW, a new ground state can be generated by an update that changes all parallel copies of the RW in the same way. Due to the periodic boundaries the stripes close to loops, as open-ended ``strings'' would incur defects of increased energy at their ends. In practice, this requires the RWs to have an even end-to-end distance (recall that $L$ must be even).  A semi-global move that again results in closed-loop stripes can be realized by flipping lines of pairs of spins along a pinned direction, which effectively inverts two consecutive steps on the RW, cf.\ Fig.~\ref{fig:groundstateMoves}. 
It is easy to see that this move locally cannot produce any open-ended ``strings'' in a ground state with a given pinned axis, and therefore this update transforms such a ground state into another with the same pinned direction. 
Note that \emph{all} ground states can be constructed using the scheme depicted in Fig.~\ref{fig:groundstateMoves}. In the top row, we have selected particularly simple configurations: In (a) the corresponding RW alternates up and down resulting in straight stripes along the horizontal armchair direction. The configuration depicted in (a) and (b) corresponds to RWs following the same direction in every step, yielding straight stripes along the vertical zigzag direction.

For axis 1 (cyan), the $L$ columns can be flipped independently yielding $2^L$ different states. In contrast, for the chosen aspect ratio each line of spin pairs along axes 2 (green) and 3 (orange) wraps the system twice, such that there are only $L/2$ independent moves in each case. Also here each such line of spin pairs can be flipped independently without changing the energy, giving rise to $2 \times 2^{L/2}$ more ground states (the factor $2$ is due to the two axes).
Note that configurations that have two pinned axes are counted twice. This is only the case for walks, which are perfectly aligned with respect to one (zigzag) axis, such as the vertical stripes shown in the top row of Fig.~\ref{fig:groundstateMoves}(b) and~(c). Similarly, stripes following only either the lattice direction 2 or 3 have two pinned axis. However, such straight-stripe configurations along axes 2 or 3 without causing defects are only possible when $L$ is divisible by 4, such that in this case six configurations are overcounted, and otherwise two.
Therefore, the total number of ground states is $2^L+2^{L/2+1}-6$ when $L$ is divisible by 4 and $2^L+2^{L/2+1}-2$ for even $L$ not divisible by 4. Note that there are exponentially more states with the vertical axis pinned, such that in simulations states with a pinning of axes 2 or 3 rarely occur and effectively can only be seen for small system sizes. 
Further, we note that above description of the ground states differs from the one given in Ref.~\cite{Azhari2025}, where the (rare) states with axis 2 or 3 pinned were overlooked~\footnote{Note that Ref.~\cite{Azhari2025} considers systems with a different aspect ratio which can, in principle, result in different degeneracies. For $L$ divisible by 4, however, one finds the same ground-state degeneracies for both aspect ratios as there is a one-to-one mapping of the ground states in this case.}.
As a result, the entropy per line is
\begin{equation}
  \frac{S(T=0)}{L}=\ln(2) + \frac 1 L \ln\!\left(1+2^{-L/2+1}-n_\text{oc}\ 2^{-L}\right),\label{eq:groundStateDegPeriodicRect}
\end{equation}
which asymptotically approaches $S(T=0)/L=\ln(2)$, in agreement with the numerical observations in Ref.~\cite{Azhari2025}. Here, $n_\text{oc}$ denotes the number of configurations counted twice, which for $L$ divisible by 4 is six, and two otherwise~\footnote{Recall that the periodic boundary conditions are only well-defined for even $L$.}. Note that the number of times lines wrap in the 2- and 3-direction depends intricately on the aspect ratio, and hence so does the exact number of ground
states. It is in general straightforward, however, to count the number of wrapping lines and hence to determine the number of independent line moves in order to obtain the correct form of Eq.~\eqref{eq:groundStateDegPeriodicRect} for such cases.

Similarly, for a system with \emph{hexagonal periodic boundaries} one arrives at $3\times 2^{L/2}-6$ ground states, that is the ground-state entropy per $L$ is
\begin{equation}
  \frac{S(T=0)}{L}=\frac 1 2 \ln 2 + \frac 1 L \ln 3 + \frac 1 L \ln\left(1-2^{-L/2+1}\right) \label{eq:groundStateDegPeriodicHex}.
\end{equation}
In each of the three (zigzag) directions there are $L/2$ lines that can be flipped independently, producing $3\times 2^{L/2}$ states from which again six have to be subtracted because of the double-counting of states without any bends, yielding Eq.~\eqref{eq:groundStateDegPeriodicHex}. (Note that for this case no $n_\mathrm{oc}$ factor is required.) Numerically, it is somewhat difficult to resolve anything but the leading contribution to the entropy. For small systems, we have carried out population annealing~(PA) simulations with very large populations to confirm Eqs.~\eqref{eq:groundStateDegPeriodicRect} and~\eqref{eq:groundStateDegPeriodicHex}; see Appendix~\ref{app:groundStateEntropy}.%

For the limiting special case $J_2=-1/4$ the degeneracy is extensive whose precise value is not known analytically. For this reason we have performed for $J_2=-1/4$ additional, especially tailored PA simulations described below in the results section Sec.~\ref{sec:minusOneQuarter}, leading to the numerical result $S(T = 0)/N = \resultZeroTEntropyMinusQuarter$ in Eq.~\eqref{eq:resultZeroTEntropyMinusQuarter}.

\begin{figure}[t]
  \includegraphics{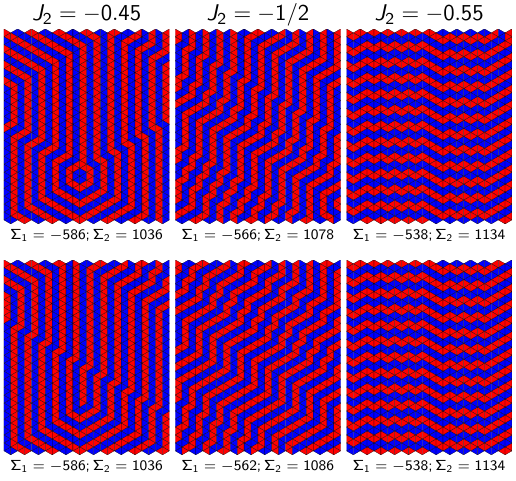}
  \caption{Exemplary snapshots of likely ground states for $L=24$ systems using rectangular FBC and for $J_2=-0.45$ (left), $-1/2$ (middle), and $-0.55$ (right). The configurations were taken from the replicas with the lowest observed energy at the final inverse temperature $\beta=5$. $\Sigma_1$ and $\Sigma_2$ refer to the \nn{} and \nnn{} energy contributions in Eq.~\eqref{eq:Hamiltonian}, respectively. Note that these states are not ground states of the periodic system.\label{fig:overviewFbcSnaps}}
\end{figure}

For \emph{free boundary conditions}, the neighborhoods of spins at the surface and the bulk are different. Thus, one cannot simply find the ground state by local energy minimization and the exact nature of the ground state will be system-size dependent.  What is more, also the exact value of the ground-state energy is no longer known.  From PA simulations of moderately-sized systems, we have randomly picked configurations of the lowest observed energy, see Fig.~\ref{fig:overviewFbcSnaps}, showing likely ground states for the $J_2$ values $-0.45$, $-1/2$, and $-0.55$, respectively. In contrast to the periodic system, the ground states are different for the selected values of $J_2$, as can easily be checked via the quoted \nn{} and \nnn{} energy contributions, which we denote by $\Sigma_1$ and $\Sigma_2$ with the Hamiltonian~\eqref{eq:Hamiltonian} reading $\mathcal{H}=J_1\Sigma_1+J_2\Sigma_2$.  One can show (see Appendix~\ref{app:energyMinFBC}), that for $-1/4>J_2>-1/2$ it is energetically favorable for the zigzag boundaries (left and right in the rectangular embedding) to be ferromagnetically aligned, and for the armchair boundaries (top and bottom) to consist of aligned pairs of spins alternating in sign.  Differently, for $J_2<-1/2$, antiferromagnetic zigzag boundaries and ferromagnetic armchair boundaries are preferred. (Note that the imperfect ferromagnetic alignment of the armchair boundaries is due to energy minimization of the corners which results in further $J_2$ dependence not discussed here.)
$J_2=-1/2$ is the marginal case for which all previously described boundaries are energetically equivalent; hence, a mixture of the scenarios is seen in the snapshots for $J_2=-1/2$. Due to the defects in the bulk, the configurations for $\rv=-0.45$ are not ground states for $\rv=-1/2$, whereas the states found using simulations for $\rv=-0.55$ have the same energy when setting $\rv=-1/2$ as the simulations for $\rv=-1/2$, but not vice-versa.

\begin{figure*}
   \includegraphics{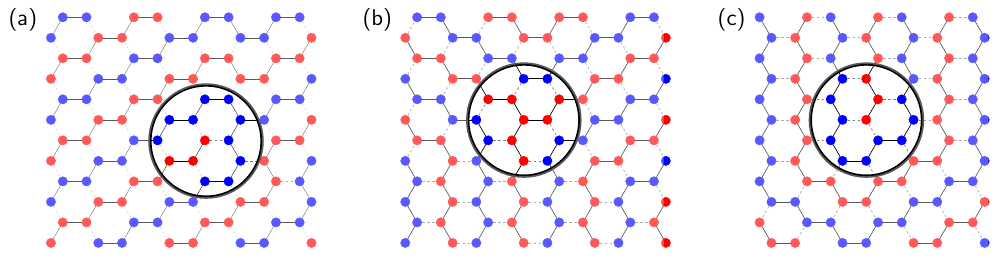}
   \caption{Exemplary representations of (a) dislocations and (b,c) disclinations that can occur. In each panel the black solid circle highlights the central part of the defect.\label{fig:topologicalDefects}}
\end{figure*}

\subsection{Topological defects}\label{sec:topologicalDefects}

As discussed above, we find that at zero temperature the threefold symmetry of preferred axes is broken by pinning one of the three directions. At sufficiently high temperature this symmetry is restored. Stripes are then parallel with respect to one of the three axes in a local region, and these regions are connected by topological defects. In close analogy to the square lattice (see, e.g., Fig.~6 of Ref.~\cite{DeBell2000}) the relevant defects are a dislocation and two types of disclinations as illustrated in Fig.~\ref{fig:topologicalDefects}.

Note that the representation in Fig.~\ref{fig:topologicalDefects} oversimplifies the actual spin configurations in two key aspects. First, the stripes locally follow one of the principal directions, whereas in the actual system they follow a RW changing between two directions with the third being pinned. Second, locally it may not be obvious which axis is pinned because only once the RW has taken two different turns at subsequent hexagons does the pinned direction become clear, which also means that boundaries between regions of different pinned axes may not be as clear as in the picture above. Nonetheless, these are the relevant defects, and they are clearly visible in actual configurations, see Fig.~\ref{fig:snapshots_-0.5} below.

\subsection{Observables}\label{sec:observables}

\begin{figure}[tb]
  \includegraphics{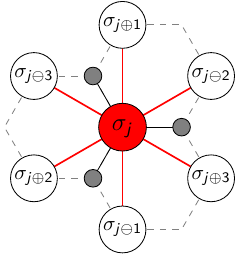}
  \caption{Illustration of how the nematic order parameter of Eq.~\eqref{eq:nematicOP} is defined.\label{fig:nematicOP}}
\end{figure}

Using the spin Hamiltonian~(\ref{eq:Hamiltonian}), energetic quantities are straightforward to measure, and we consider the energy per spin, $e = \langle\mathcal{H}\rangle / N$, and the specific heat, $C_V = (\langle\mathcal{H}^2\rangle - \langle\mathcal{H})\rangle^2 / N T^2$.
On the other hand, to date no order parameter is known for this model that could distinguish the stripe state at zero temperature from the high-temperature paramagnetic phase.
To fill this gap, in analogy to nematic liquid crystals~\cite{Chandrasekhar2004}, we define a local nematic order parameter $\eta_j$ as
\begin{equation}
  \eta_j = \sum_{k=1}^3 \frac 1 2 (\sigma_j \sigma_{j\oplus k} + \sigma_j \sigma_{j\ominus k}) f_{k}, \label{eq:nematicOP}
\end{equation}
where the sum is over the three principal lattice directions, $(0,1)$ ($k=1$), $(-\sqrt{3}/2,-1/2)$ ($k=2$), and $(\sqrt{3}/2,-1/2)$ ($k=3$). The spin $\sigma_{j\oplus k}$ ($\sigma_{j\ominus k}$) refers to the next-nearest neighbor of $\sigma_j$ in positive (negative) $k$ direction; see Fig.~\ref{fig:nematicOP} for an illustration. $f_{1} = 1, f_2 = e^{2\pi i / 3}$, and $f_3 = e^{4\pi i / 3}$ encode the direction as a complex number~\footnote{Note that this is a generalization of using $\pm 1$ for stripes on a square lattice as is done, e.g., in Ref.~\cite{Ye2021}. An equivalent formulation in vector notation for nearest neighbors on a triangular lattice can be found in Ref.~\cite{Simmons2024}.}.

$\eta_j$ acts as a local order parameter: When stripes locally are oriented in one of the principal directions, $\eta_j$ takes values $2e^{2\pi i / 3}, 2e^{4\pi i / 3}$, or $2$, that is, its absolute value is two and its argument $\phi = 2\pi k/3$ with $k=1,2,3$ corresponds to the direction the stripes follow. Thus, if stripes were globally ordered along one of the lines, we would expect $\langle |\eta| \rangle = 2$. Here, $\eta$ refers to the mean value of $\eta$ per site, i.e., $\eta = (\sum_j \eta_j)/N$. On the other hand, when all directions occur with equal frequencies, $\langle |\eta| \rangle = 0$ in the thermodynamic limit. However, as discussed above, the meandering stripes in the low-temperature phase do not have one preferred direction but instead one direction is pinned, and they follow RWs along the two remaining directions. Excluding one direction, yields $\langle |\eta| \rangle = 1$ (see Appendix~\ref{app:etaZeroTemp}).

Further, we define the nematic susceptibility as
\begin{equation}
  \chi = N \left(\langle |\eta|^2 \rangle - \langle |\eta| \rangle^2\right),\label{eq:susceptibility}
\end{equation}
and the fourth-order nematic Binder parameter~\cite{binder:81,Viet2009,Tuan2022} is given by
\begin{equation}
   U_4 = 2 - \frac{\langle |\eta|^4 \rangle}{\langle |\eta|^2 \rangle^2}\label{eq:BinderCumulant}, %
\end{equation}
which as in Refs.~\cite{Viet2009,Tuan2022} is normalized such that it is zero at infinite temperature and one at zero temperature.

\section{Algorithm and simulation details}\label{sec:algo}
\subsection{Population annealing}\label{sec:algoPA}

PA is an algorithmic framework in which a population of replicas is sequentially cooled down using independent Markov chain Monte Carlo (MCMC) moves at each temperature, followed by a population control step. The population of $R_0 = R$ replicas is initialized by randomly drawing a spin configuration for each replica at the initial inverse temperature $\beta_0$. Here, $\beta_0=0$ (infinite temperature) is a convenient choice, as every spin can be picked at random $+1$ or $-1$. Having initialized the population, and setting the annealing index $i$ to zero, the algorithm proceeds as follows: \begin{enumerate}[itemsep=2pt] \item[(i)] For every replica $k$ calculate \begin{equation} \tau_{i}(E_k) = \frac{R_0}{R_i} e^{-\Delta\beta E_k} / Q_i,\label{eq:PA_tau} \end{equation} where $\Delta\beta\coloneqq\beta_{i+1}-\beta_i$ is the inverse temperature step, $E_k$ the energy of replica $k$, $R_i$ the population size at $\beta_i$, and $Q_i = \sum_{j=1}^{R_i} e^{-\Delta\beta E_j}$.  \item[(ii)] Resample every replica according to Eq.~\eqref{eq:PA_tau}, i.e., replica $k$ should on average appear $\tau_i(E_k)$ many times in the resampled population. Increment $i$ by 1.  \item[(iii)] Carry out $\theta_i$ rounds of Metropolis updates.  \item[(iv)] Measure estimates for observables $\mathcal{O}$ via \begin{equation} \langle \mathcal {O} \rangle_{\beta_i} \approx \hat{\mathcal{O}} = \frac{1}{R_i} \sum_{k=1}^{R_i} \mathcal{O}_k, \label{eq:popAverage} \end{equation} where $\langle \dots \rangle_{\beta_i}$ denotes a thermal average at inverse temperature $\beta_i$.  \item[(v)] Repeat steps (i)-(iv) until the final inverse temperature $\beta_\text{f}=1/T_\text{f}$ is reached.  \end{enumerate} This general framework still leaves much room to adjust the various parameters, in particular the population size $R$, the set of inverse temperatures $\{\beta_i\}$ (the so-called annealing schedule), the number of Monte Carlo sweeps (MCS) per replica $\{\theta_i\}$ in each iteration, and which resampling method to use. In the simulations reported on here, we use the following recipes. For the annealing schedule, the temperatures are selected adaptively~\cite{Barash2017}: Before carrying out step~(i) we determine a step size $\Delta\beta$ for which the histogram overlap \begin{equation} \alpha(\Delta\beta) = \frac{1}{R_i} \sum_{k=1}^{R_i} \min\left(1, \frac{R}{R_i} \frac{\exp(-\Delta\beta E_k)}{\sum_{j=1}^{R_i} e^{-\Delta\beta E_j}}\right) \end{equation} is close to a target value of $\alpha^{*}$ by means of numerical root finding; this $\Delta\beta$ is then used in step~(i).

Similarly, for the rounds of MCS, $\theta_i$ is selected adaptively by carrying out sweeps until an equilibration criterion is met~\cite{Gessert2025}. For the criterion the \emph{effective population size} $R_\text{eff}$~\cite{Weigel2021} is used, which is analogous to the effective sample size $N_\text{eff} = N / 2 \tau_{\mathcal{O},\text{int}}$ for a correlated time series of an observable $\mathcal{O}$ with the integrated autocorrelation time $\tau_{\mathcal{O},\text{int}}$. Specifically, for the adaptive sweep schedule, MCS are performed until $R_\text{eff} / R$ is larger than its target value $\rho^{*}$. In practice, we use a heuristic cut-off to ascertain simulations reach the final temperature (see Sec.~\ref{sec:simulationDetail} for details).

The resampling step (ii) is realized by the nearest-integer resampling method~\cite{Gessert2023} in which given $\tau$ the number of descendants $r$ follows the probability mass function $P_\tau(r)$ given by \begin{equation} P_{\tau}(r) = \begin{cases} \tau - \lfloor\tau \rfloor & \text{if } r = \lfloor\tau \rfloor + 1 \\ 1 - (\tau - \lfloor\tau \rfloor) & \text{if } r = \lfloor\tau \rfloor \\ 0 & \text{else} \end{cases}, \end{equation} where $\lfloor \tau \rfloor$ refers to the largest integer smaller than or equal to $\tau$.

\subsection{Simulation details}\label{sec:simulationDetail}

The computer programs for our simulations are based on the publicly available code of Ref.~\cite{Barash2017}, which implements the PA algorithm on GPUs using CUDA. Besides the per-replica parallelism, the implementation makes use of domain decomposition of the spin configurations for the Metropolis updates providing a sub-replica parallelism. The domain decomposition taking into account the nearest and next-nearest neighbors divides the lattice into eight sublattices~\footnote{The corresponding coloring problem can be solved with four colors. However, we opt to assigning twice as many spins to each thread as it simplifies the thread-spin mapping and also with half as many parallel threads the GPU is fully utilized.}. Further speed-up could potentially be achieved by using the multi-spin coding version of the code. However, our simulations used the single-spin coding version of Ref.~\cite{Barash2017}.

All simulations use a population of $R = 20\,000$ replicas. The target overlap of the energy histograms at neighboring inverse temperatures is $\alpha^{*}=0.8$, and the effective population size threshold for equilibration is set to $\rho^{*}=0.9$. To avoid having to calculate the effective population size too frequently and to avoid failing to reach the final inverse temperature by never satisfying the equilibration criterion, we use the following heuristic: $\Delta \theta$ MCS are carried out before calculating $R_\text{eff}$. The value for $\Delta \theta$ at the beginning of the simulation is 100 MCS. At each new temperature, $\Delta \theta \leftarrow \Delta \theta / 2$. When after $\Delta \theta$ MCS $R_\text{eff}/R$ is less than $\rho^{*}$, set $\Delta \theta \leftarrow 2\Delta \theta$, and carry out further $\Delta \theta$~MCS. An exception to this case is when $\theta_i+\Delta\theta$ exceeds \texttt{max\_sweeps}, with $\theta_i$ being the sweeps already carried out and \texttt{max\_sweeps} initially set to 5000 MCS. In this case, $\Delta \theta \leftarrow \texttt{max\_sweeps} -\theta_i$. If after \texttt{max\_sweeps}~MCS $R_\text{eff}/R$ is still less than $\rho^{*}$, the simulation proceeds to the next inverse temperature unless $R_\text{eff}/R$ is less than $0.4$. In this case, \texttt{max\_sweeps} is doubled and one proceeds as above. Finally, the last cut-off is $\texttt{max\_max\_sweeps} = 5\times10^5$~MCS which neither $\theta_i$ nor \texttt{max\_sweeps} is allowed to exceed. When \texttt{max\_max\_sweeps} is reached, the simulation proceeds with the next annealing step.

As for the representation of the Ising spin configurations on the honeycomb lattice, they are encoded using two-dimensional arrays. Rectangular systems of linear size $L$ and employing PBC or FBC contain $L^2$ hexagons comprised of $N=2 L^2$ spins, and they are represented by an $L_x \times L_y$ array with $L_x=2L$ and $L_y=L$. For an illustration how exactly configurations translate from physical coordinates to ``memory'' coordinates, we refer to Appendix~\ref{app:representationHoneycomb}, also showing how this representation relates to the frequently used brick-wall representation. PBCs are implemented using the same modulus operations one would use on a square lattice. Less straightforward is the implementation of the hexagonal boundaries. Here we opt for using a neighbor table which, however, substantially reduces the simulation performance of our GPU code and hence the system sizes that can be studied.

\section{Results}\label{sec:results}

From our PA simulations we extracted the behavior at three different NNN couplings: $J_2 = -1/4$ at the threshold between the striped ground states and the ferromagnetic ones, and $J_2 = -0.5$ as well as $J_2 = -1$ within the partially-ordered stripe phase. After presenting the corresponding results, we will discuss the sensitive dependence on aspect ratios alluded to above.

\subsection{$J_2=-1/4$}\label{sec:minusOneQuarter}
\begin{figure*}[ht]
  \includegraphics{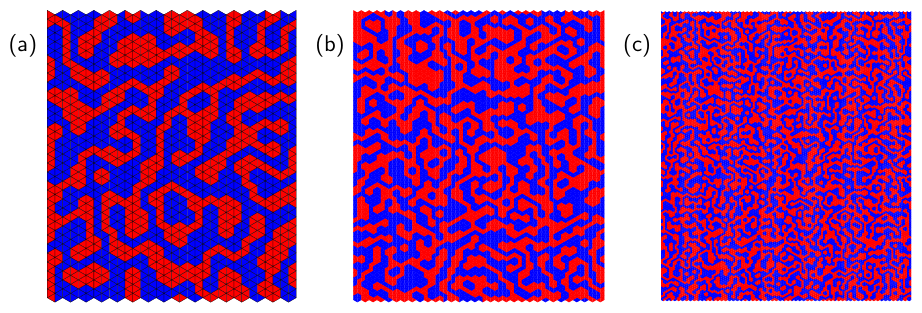}
  \caption{Example ground-state spin configurations for the special point $\rv=-1/4$ and different system sizes $L$. (a) $L=32$, (b) $L=64$, and (c) $L=128$. In panel (a) the underlying hexagonal lattice is highlighted by solid black lines.\label{fig:snaps_J2_-0.25}}
\end{figure*}

At the special point $\rv= -1/4$, both the ferromagnetic state (comprising of all spins pointing up or all spins pointing down) and the highly degenerate stripe state have the same energy, which results in an even more degenerate ground state at this special point. Thermodynamically, the transition temperature is expected to vanish for $\rv=-1/4$~\cite{Bobak2016,Schmidt2021,Zukovic2021,Azhari2025,Batista2026}, and the ground-state entropy per spin should remain finite at zero temperature~\cite{Azhari2025,Batista2026}.

We first consider typical ground-state configurations as they occur in the simulations. (Note that identifying ground states is straightforward since the ground-state energy is exactly known.) The examples shown in Fig.~\ref{fig:snaps_J2_-0.25} reveal structures that resemble a kind of stripe liquid with some degree short-range correlations --- somewhat similar to the hexagonal phase seen in the dipolar model on the honeycomb lattice~\cite{Rueger2012}. From the snapshots for different system sizes shown in Fig.~\ref{fig:snaps_J2_-0.25} it is clear that these ground states do not exhibit conventional long-range order.

\begin{figure}[tb!]
  \includegraphics{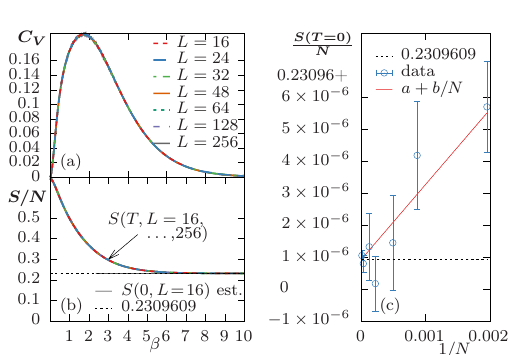}
  \caption{(a) Specific heat $C_V$ and (b) entropy per spin $S/N$ vs.\ $\beta$ for different system sizes $L$ and $\rv=-1/4$. (c) shows the ground-state entropy estimate for different system sizes resulting from Eq.~\eqref{eq:pa_S_GS_estimator} at the lowest temperature $T_\text{f}=0.1$ (except for $L=32$ where $T_\text{f}=0.05$). The dashed lines in (b) and (c) correspond to the best estimate for $S(T=0)$, and the almost horizontal solid black line in (b) shows a numerical evaluation of the ground-state entropy estimator~\eqref{eq:pa_S_GS_estimator} for $L=16$ as a function of $\beta$.\label{fig:Cv_and_S_J2quarter}}
\end{figure}

To arrive at a more quantitative description, we consider the specific heat $C_V$ and the entropy per site $S/N$ as a function of inverse temperature $\beta$ as shown in Figs.~\ref{fig:Cv_and_S_J2quarter}(a) and~(b). The specific-heat curves in Fig.~\ref{fig:Cv_and_S_J2quarter}(a) are rather featureless, and show no system-size dependence such that the data for all sizes from $L=16$ to $L=256$ practically lie on top of each other~\footnote{Note that here we consider larger system sizes and use larger population sizes as compared to those of Refs.~\cite{Gessert2024,Gessert2025} and the simulations in the stripe phase discussed below. Due to the absence of critical slowing down, simulations are very easy to equilibrate at $\rv = -1/4$.}. This is consistent with similar observations in Refs.~\cite{Azhari2025,Batista2026} using Monte Carlo simulations of smaller systems and a correlated cluster mean-field (CCMF) approach, respectively.  The entropy per site also shows no system-size dependence; starting from the high-temperature limit of $\ln 2 \approx 0.69$ at $\beta=0$ it smoothly approaches a zero-temperature value of approximately $0.23$.  This further indicates that for $\rv=-1/4$ the model is thermodynamically trivial.  Note that this value differs significantly from the approximation of $0.499$ within the CCMF approach of Ref.~\cite{Batista2026}.

Although thermodynamically trivial, no exact solution is known for $\rv= -1/4$ (in contrast to the case $\rv \rightarrow -\infty$) and even the ground-state entropy per spin is unknown.  Before determining the residual entropy per spin numerically to high precision, we present a simple argument as to why it is non-zero in the first place: Starting from a fully ordered state (which also minimizes the energy for $\rv=-1/4$), a single hexagon of spins can be flipped at no energetic cost~\cite{Gessert2025}. Clearly, the same is true for any other hexagon not sharing any links with the first one, and as there are only short-range interactions, the number of non-interacting hexagons is extensive. The orientation of the spins on each such hexagon can be arbitrarily and independently chosen up or down, yielding at least $2^{O(N)}$ ground-state configurations, and hence $\ln \Omega(E_\text{GS})$ scales linearly in $N$. While this shows the existence of residual entropy in this model, clearly the constructed configurations have little resemblance with those of Fig.~\ref{fig:snaps_J2_-0.25}. In fact, since there are $N/8$ non-interacting hexagons on a hexagonal lattice with $N$ spins, this description corresponds to $\sim 2^{N/8}$ ground states, only providing a crude lower bound for the ground-state entropy per spin of $\ln (2) /8 \approx 0.0866 \ll 0.23$.

Next, we turn to numerically estimating the residual, ground-state entropy.  The entropy at positive temperatures can be quite conveniently estimated using population annealing~\cite{Barash2017,barash:18}, such that it appears to be a natural choice also for determining $S(T=0)$. However, one usually stops the annealing before reaching zero temperature, and thus the ground-state entropy cannot be directly measured in PA.
As we show in Ref.~\cite{Gessert2026}, $S(T=0)$ can be estimated at non-zero simulation temperatures $T$ using
\begin{equation}
   S(T=0) = S(T) + \ln[\rho_\text{GS}(T)] - \frac{E(T)-E_\text{GS}}{T}, \label{eq:pa_S_GS_estimator}
\end{equation}
where $S(T)$ and $E(T)$ are the (extensive) entropy and energy, respectively, at the simulation temperature $T$, $\rho_\text{GS}(T)$ is the fraction of the population that has reached the ground state, and $E_\text{GS}$ is the ground-state energy. One should note that the estimator~\eqref{eq:pa_S_GS_estimator} can only be evaluated when the estimate for $\rho_\text{GS}(T)$ is larger than zero.

In Fig.~\ref{fig:Cv_and_S_J2quarter}(b), we display the entropy for different system sizes $L$, again showing practically no $L$ dependence as was the case for $C_V$.  In addition to the PA estimate for the entropy $S(T)$, we also show the estimates for the ground-state entropy obtained from Eq.~\eqref{eq:pa_S_GS_estimator} for system size $L=16$ as a function of the inverse temperature $\beta$ used in the simulation (almost horizontal solid black line). Since Eq.~\eqref{eq:pa_S_GS_estimator} cannot be evaluated when no replica has reached a ground state yet, i.e., when the estimate for $\rho_\text{GS}$ is zero, the line does not extend all the way down to $\beta=0$.  On the scale of the plot the estimate of Eq.~\eqref{eq:pa_S_GS_estimator} for $L=16$ is indistinguishable from the final, finite-size extrapolated estimate for $S(T=0)$ shown as a dashed line.
The estimate of Eq.~\eqref{eq:pa_S_GS_estimator} at the lowest simulated temperature is used as final value of $S(T=0)$ for each system size in Fig.~\ref{fig:Cv_and_S_J2quarter}(c).  The asymptotic residual entropy per site is obtained via a linear fit in $1/N$ yielding the estimate 
\begin{equation}
S(T=0)/N = \resultZeroTEntropyMinusQuarter.\label{eq:resultZeroTEntropyMinusQuarter}
\end{equation}
(The fit with five degrees of freedom had a reduced $\chi^2$ of $0.8$.) Note that such a $1/N$ correction appears naturally assuming the number of ground states (to leading order) is $\tilde{b}\times \tilde{a}^N$, where $\tilde{a}$ and $\tilde{b}$ can be identified as $\tilde{a}=\ln a$ and $\tilde{b}=\ln b$ of the fitting function $a+b/N$ in Fig.~\ref{fig:Cv_and_S_J2quarter}(c).

\subsection{$J_2=-0.5$}\label{sec:minusOneHalf}

We now move into the partially-ordered stripe phase, first presenting results for $\rv = -0.5$. In introducing the results for $\rv < -1/4$ for the time being we refrain from interpreting them in terms of the presence or absence of a phase transition of a specific type. We postpone this discussion until Sec.~\ref{sec:discussion} when all data have been introduced.

In previous work~\cite{Zukovic2020} it was observed that for $\rv = -0.5$ the model is easier to equilibrate than for values both above and below $-0.5$, where dynamics reminiscent of the slow relaxation typical of spin glasses has been reported~\cite{Zukovic2020,Zukovic2022}.  Except for $L=128$ using (rectangular) PBC and $L=160$ using (rectangular) FBC, all data presented here was averaged over at least 10 independent PA runs, which also provides a convenient means for obtaining the standard error. The annealing schedule was determined in the first run, and then kept fixed in the other runs of the same system size.

\begin{figure}[tb!] 
  \includegraphics{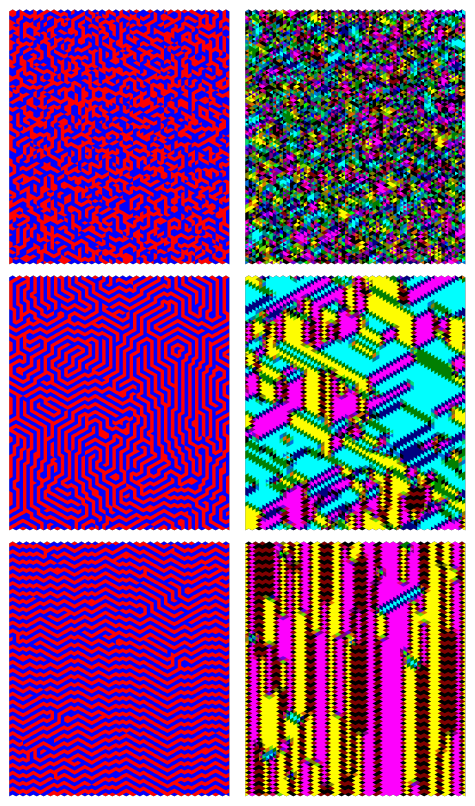}
  \caption{Selected snapshots for $\rv = -0.5$ with rectangular PBC and $L=64$ for different inverse temperatures $\beta$, namely $\beta = 0.80$ (top row), $\beta = 2.1$ (middle row), and $\beta = 2.31$ (bottom row). The left column shows the Ising spin configurations, and the right column the local nematic order parameter $\eta_j$ from Eq.~(\ref{eq:nematicOP}) encoded using the HSV color representation with the absolute value of $\eta_j$ corresponding to the value V and the complex angle as hue H.\label{fig:snapshots_-0.5}}
\end{figure}

In Fig.~\ref{fig:snapshots_-0.5}, we present three representative snapshots of spin configurations with (rectangular) PBC and $L = 64$ at three different temperatures. The left column shows the actual Ising spin configurations $\sigma_j$, and the right column shows the local values of the complex nematic order parameter $\eta_j$ defined in Eq.~\eqref{eq:nematicOP} with its argument encoded by the color hue, and its absolute value by how bright or dark it is. The temperatures are selected such that the first ($\beta = 0.80$, top row) is well above the potential transition temperature, the second ($\beta = 2.1$, middle row) close to the transition temperature, and the third ($\beta = 2.31$, bottom row) well within the stripe phase. In the latter case the vertical direction is pinned.  For the high-temperature configuration, no order is visible, whereas in the middle row some sort of stripe order appears. This is more clearly seen when considering the nematic order parameter configurations: Where domains of three different colors meet in the $\eta$-representation, topological defects can be seen in the spin configuration, cf.~Fig.~\ref{fig:topologicalDefects}. The stripes in the spin configuration of the bottom panel are almost fully aligned according to the pinned vertical axis. The stripes then randomly meander along the two remaining directions. This is seen in the $\eta$-representation by the colors purple and yellow appearing at random. The red and black stripe pattern seen, e.g., in the top-left corner, appears when locally the purple and yellow directions are `mixed'. Due to the pinning of the vertical axis, the color cyan is almost entirely absent in this representation, which is reflected in the spin configuration only showing very few defects.

It is interesting to note that the spin configurations we observe in the vicinity of the potential transition temperature closely resemble the spin configurations observed in the dipolar Ising model on the triangular lattice with a purely dipolar interaction term~\cite{Vedmedenko1998}, as well as in a generalized version which includes a six-state clock model interaction~\cite{Simmons2024}; see Fig.~2 in Ref.~\cite{Vedmedenko1998} and Fig.~5(a2) in Ref.~\cite{Simmons2024}. 
In fact, Ref.~\cite{Simmons2024} considers an order parameter analogous to the one introduced here, albeit using a vector notation rather than complex numbers.

As discussed in Sec.~\ref{sec:groundStates}, the $\eta$-symmetry-breaking related to the preferred pinning of the vertical axis is induced by the rectangular PBC in which pinning in one of the remaining two directions is exponentially suppressed.  This is also reflected in the percolation properties: At some intermediate temperature above the potential transition temperature the striped structures, e.g., in the middle row of Fig.~\ref{fig:snapshots_-0.5}, are likely to percolate in both the vertical and the horizontal direction. Once the vertical axis is pinned, stripes avoid the vertical direction which is reflected in a sudden drop of the vertical percolation probability while leaving the one in the horizontal direction unchanged; see Appendix~\ref{app:percolation} for more details.  This further motivates studying multiple types of boundary conditions.
In addition, we note that the aspect ratio for a rectangular embedding also affects observations, and different ratios were used in the literature; see Sec.~\ref{sec:aspectRatio}.  To simplify the discussion below, however, we only report results for our aspect ratio~\cite{Gessert2024,Gessert2025} in the main text, i.e., $L_x:L_y = 2\!:\!1$.

\begin{figure*}[tb!]
  \includegraphics{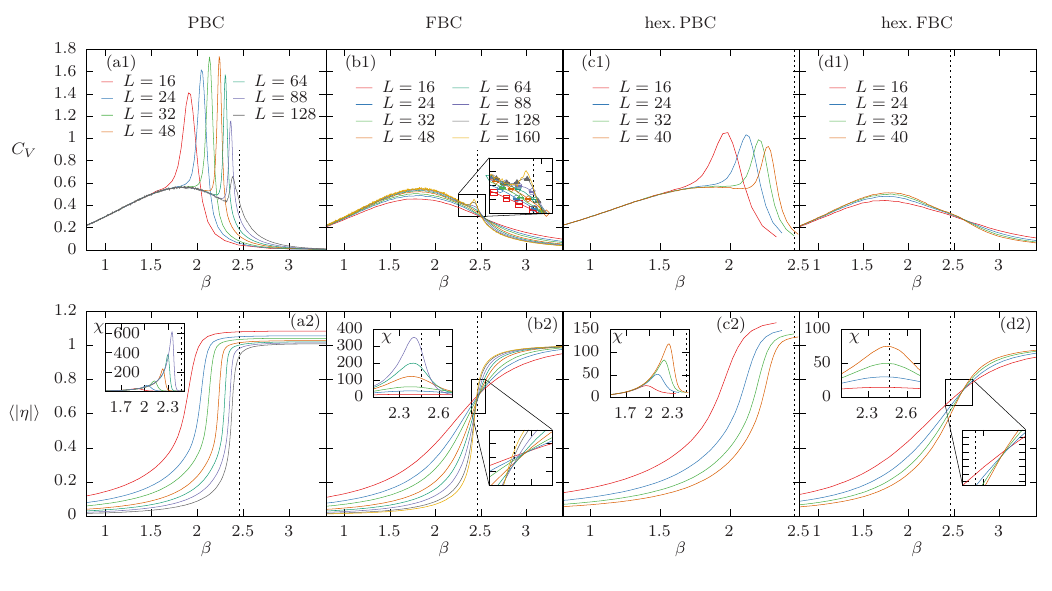}
  \caption{Specific heat (top) and mean absolute value of the nematic order parameter (bottom) for different types of free and periodic boundary conditions, and using $J_2 = -0.5$. The insets in the bottom row show the susceptibility $\chi$ given by Eq.~(\ref{eq:susceptibility}). In all plots and insets the vertical dashed line corresponds to $\beta = 2.46$.\label{fig:overview_-0.5}}
\end{figure*}

The top row of Fig.~\ref{fig:overview_-0.5} shows the specific heat as a function of inverse temperature $\beta$ for different system sizes and different types of boundary conditions. Here, panel~(a1) corresponds to the previously studied case~\cite{Zukovic2020,Azhari2025} (but note the different aspect ratios employed in these works).
The most immediate observation when considering the different types of boundaries is that the sharp peak in the specific heat, whose location is strongly system-size dependent, is not present when using free boundary conditions. Only for the largest system sizes considered a small peak appears to develop, and its location appears to be much less dependent on $L$ (see zoomed-in region in the inset). We also notice that the two types of hexagonal boundary conditions behave quite similar to their rectangular counterparts, such that we will focus the finite-size scaling discussion on the rectangular boundaries for which we can simulate substantially larger systems. For free boundary conditions, the specific-heat curves appear to cross at an inverse temperature close to $2.5$. However, a closer look reveals that this only occurs for smaller system sizes and for $L \geq 88$ the apparent crossing moves further towards lower $\beta$.

\begin{figure}[tb!]
  \includegraphics{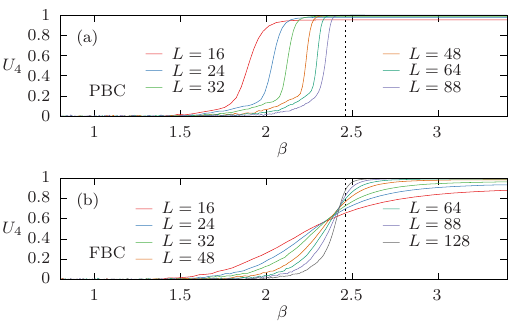}
  \includegraphics{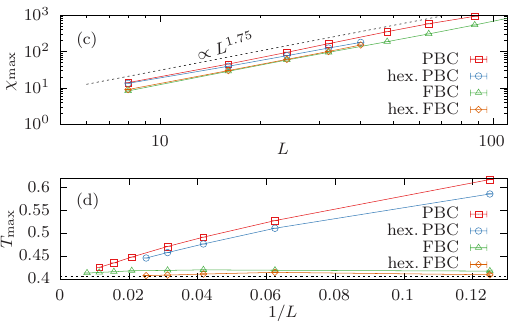}
  \caption{Finite-size scaling plots for $J_2=-0.5$. The Binder parameter for different $L$ using rectangular (a) PBC and (b) FBC. The peak heights of the susceptibility are shown in panel (c), and the peak locations $T_{\max}$ in (d). The horizontal dashed line in (d) corresponds to the potential transition temperature $T = 1 / 2.46 \approx 0.407$. \label{fig:fss_-0.5}}
\end{figure}

The bottom panel of Fig.~\ref{fig:overview_-0.5} shows the mean absolute value of the nematic order parameter $\langle|\eta|\rangle$ as well as its respective susceptibility $\chi$ in the insets. At high temperature, $\langle|\eta|\rangle$ is close to zero, and it takes a non-zero value close to one at low temperatures (see Appendix~\ref{app:etaZeroTemp} for a discussion of the zero-temperature limit of $\langle|\eta|\rangle$).
For both types of periodic boundary conditions, the increase in $\langle|\eta|\rangle$ aligns with the specific-heat peak, and it also is strongly system-size dependent. The curves for $\langle|\eta|\rangle$ appear to cross at some temperature for free boundary conditions, albeit at a different temperature for the two types of free boundary conditions, and also at a different temperature than the apparent crossing of the specific heat. Also here, a closer look at the largest system sizes reveals that the apparent crossing is only visible in smaller system sizes, and for larger systems crossings of two consecutive sizes shifts towards lower $\langle|\eta|\rangle$ and lower~$\beta$. The vertical line $\beta = 2.46$ is a guide to the eye that is drawn to facilitate the comparison of the different plots of Fig.~\ref{fig:overview_-0.5}. Also, with the limited numerical data available, the peaks in $C_V$ and~$\chi$, as well as the location of the increase in $\langle|\eta|\rangle$ are consistent with approaching $\beta = 2.46$ as $L \rightarrow \infty$.

Next, we carry out a finite-size scaling analysis using the temperature behavior of the nematic Binder parameter, as well as the scaling of the susceptibility for the different boundary conditions, see Fig.~\ref{fig:fss_-0.5}. For periodic boundary conditions, no crossing is visible in the Binder parameter, panel (a), whereas for free boundary conditions, a crossing for different $L$ is seen, %
although crossings of larger $L$ shift towards higher values of $\beta$.

The peak heights of the susceptibility grow according to a power-law with an exponent of $1.75$ for the four types of boundaries, cf.\ Fig.~\ref{fig:fss_-0.5}(c). The behavior of the peak locations, on the other hand, strongly depends on the boundary conditions. As was the case for the specific heat, the peak locations vary strongly with system size for both types of periodic boundary conditions, and show little system-size dependence for free boundaries. In all four cases, the $T_{\max}$ vs.\ $1/L$ plot is strongly suggestive of the temperatures converging to a non-zero value as $L\rightarrow \infty$. (Here, we show temperature $T$ rather than $\beta$ because visually it is easier to discern whether $T$ approaches zero or not than whether $\beta$ diverges --- very slowly --- or approaches a constant.)  Note that if one were to assume a power-law ansatz $|T_{\max} - T_t| \propto L^{-1/\nu}$, the less-than-linear growth of $T_{\max} (L)$ would translate to $\nu > 1$, which using the hyperscaling relation $\nu d = 2 - \alpha$ implies a negative $\alpha$. This agrees with Ref.~\cite{Azhari2025} also reporting a value of $\nu$ larger than one. 
Negative $\alpha$ translate to a non-divergent specific heat, which is consistent with the heights of the specific-heat peaks seen in Fig.~\ref{fig:overview_-0.5}(a1)-(d1).

\subsection{$J_2=-1$}\label{sec:minusOne}

\begin{figure*}[tb!]
  \includegraphics{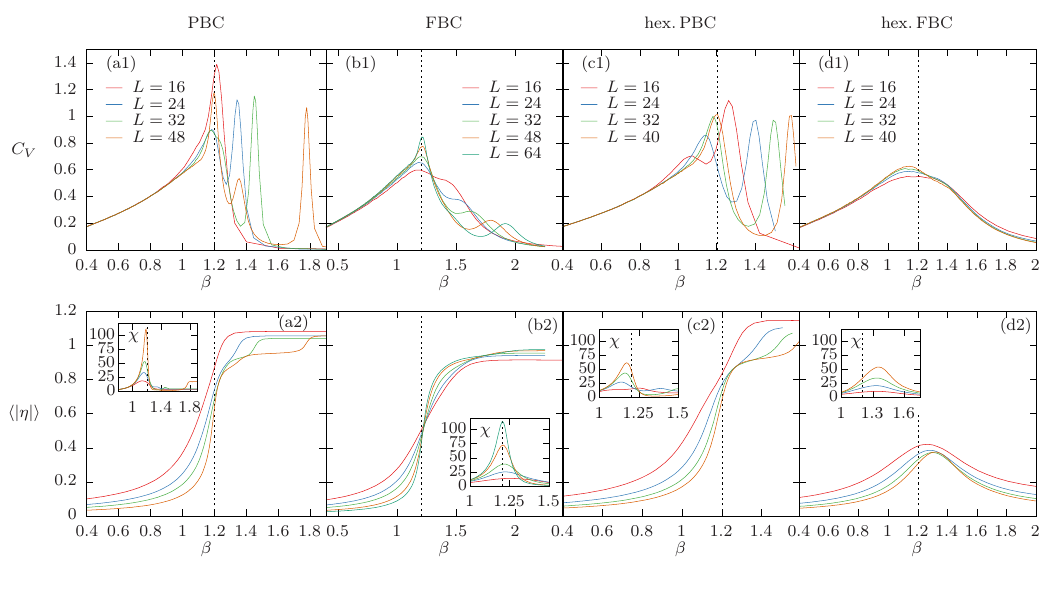}
  \caption{As Fig.~\ref{fig:overview_-0.5}, but using $J_2 = -1$.\label{fig:overview_-1}}
\end{figure*}

We now turn to the more strongly frustrated case of $J_2 = -1$. This coupling ratio has been studied in Ref.~\cite{Zukovic2022} using parallel tempering Monte Carlo, reporting on features in the specific heat whose number appeared to increase with system size. From the scaling of $C_V$ and inspection of the energy histograms the author conjectured the transition at $\beta \approx 1.2$ to be of continuous type, and the transitions for larger $\beta$ to have first-order character (based on the bimodal nature of the energy histograms in Ref.~\cite{Zukovic2022}). Figure~\ref{fig:overview_-1} shows the specific heat and the nematic order parameter as a function of inverse temperature for the different types of boundary conditions. For PBC, the locations of the specific heat peaks in Fig.~\ref{fig:overview_-1}(a1) are in good agreement with those of Ref.~\cite{Zukovic2022}. The system sizes in common are $L=24$ and $L=48$. Note, however, the different used aspect ratio which in this case affects predominantly the height of the specific-heat peaks, and only slightly their locations; see Sec.~\ref{sec:aspectRatio} for a discussion of the effect of different aspect ratios in this model.

Regarding the potential transition near $\beta = 1.2$ (vertical dashed lines), with the exception of $L=16$, the $C_V$ peak grows with system size for PBC. For $L=16$ the high-temperature peak merges with the peak at the next-lower temperature, leading to the larger peak. This is more clearly visible in Ref.~\cite{Zukovic2022} through the inclusion of data for $L=72$. Also for the other boundary conditions, the specific heat shows a relatively clear signal in the vicinity of $\beta = 1.2$. At the same inverse temperature, $\langle|\eta|\rangle$ shows an increase from close-to-zero to a non-zero value. For both rectangular and hexagonal PBCs, this non-zero value is different from the zero-temperature value, and a plateau for intermediate temperatures is visible. This plateau is absent in rectangular FBC for which $\langle|\eta|\rangle$ appears to directly approach its zero-temperature value.
Finally, for hexagonal FBC $\langle|\eta|\rangle$ has a maximum near $\beta = 1.2$ and appears to take a value close to zero for low temperatures. This effect is understood by considering the nature of the boundaries, all of which in this case are zigzag edges [cf. Fig.~\ref{fig:honeycombLattice}(b)], which for $J_2=-1$ favor to be antiferromagnetically aligned (see Appendix~\ref{app:energyMinFBC}). As there are two zigzag boundaries along each of the three principal directions, the threefold symmetry is effectively enforced by the boundary conditions, at least for the small systems considered here, giving rise to the maximum in $\langle|\eta|\rangle$ and its low values at low temperature. This effect was not seen in Fig.~\ref{fig:overview_-0.5}(d2) because for $J_2 = -1/2$, the zigzag boundaries provide no energetic preference for either ferromagnetic or antiferromagnetic order, and therefore they do not reinforce the threefold symmetry at low temperatures in that case.
More generally, the threefold symmetry at low temperatures is expected for any $J_2 > -1/2$ as also ferromagnetically aligned zigzag boundaries found there enforce this symmetry. Preliminary Monte Carlo simulations of values of $J_2$ slightly above or below $-1/2$ (not shown) indeed indicate that $J_2=-1/2$ is the exception, and that for $J_2 \neq -1/2$, $\langle|\eta|\rangle$ is small at low temperatures when using hexagonal FBCs at least for the relatively small system sizes considered.

Except when employing hexagonal FBC, numerous features are visible in $C_V$ and $\langle|\eta|\rangle$, and their number appears to increase with system size.
All such features have strongly system-size dependent locations (moving rapidly towards lower temperatures with increasing $L$), that additionally heavily depend on boundary conditions. Peaks in $C_V$ appear to be linked with sharp increases in $\langle|\eta|\rangle$ at the same temperature.  The sharp rises in $\langle|\eta|\rangle$ are separated by plateaus.  When two $C_V$ peaks are in close proximity, however, the plateau $\langle|\eta|\rangle$ is only visible in form of an inflection point, see, e.g., the $C_V$ peaks at $\beta \approx 1.2$ and $1.35$ for $L=24$ and $L=48$.

\begin{figure}[tb!]
  \includegraphics{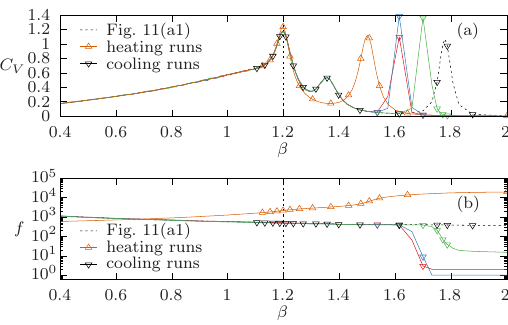}
  \caption{Heating and cooling runs for $L=48$, $J_2=-1$, and using rectangular PBC. Cooling runs shown in different colors are individual runs that experienced a sudden jump towards an energy close to the ground state (see text). The heating runs are averaged over 10 realizations. (a) Specific heat and (b) the number of surviving families. The dashed lines show the data from Fig.~\ref{fig:overview_-1}(a1).\label{fig:takeover_-1_L48}}
\end{figure}

Based on the bimodality of the energy histograms at the respective temperatures, in Ref.~\cite{Zukovic2022} these features were described as being first-order-like. Despite the bimodal character of the histograms hinting at a possible first-order transition, the height of the specific-heat peak is almost independent of system size. Inspection of the snapshots using rectangular PBC below and above the peak at $\beta\approx 1.8$ for $L=48$ show that for $\beta < 1.8$ regions where stripes follow the vertical direction do exist whereas they are absent for $\beta > 1.8$. While most simulations for this system size behaved as in Fig.~\ref{fig:overview_-1}(a1) and (a2), some runs showed a sudden jump in $\langle|\eta|\rangle$ as well as a sharp $C_V$ peak at lower $\beta$. Further, those runs showed clear signs of failed equilibration, and we therefore excluded them in Fig.~\ref{fig:overview_-1}(a1) and (a2).
However, this also raises concerns about the reliability of the low-temperature features when runs did appear to be equilibrated. Ultimately, PA in its canonical form relies on sufficient histogram overlap between successive temperatures, and therefore is not well suited for first-order transitions~\cite{Barash2017a}. Hence, in order to better understand these features, methods more suitable for the first-order-like behavior should be used.

Figure~\ref{fig:takeover_-1_L48} shows (a) the specific heat $C_V$ and (b) the number of surviving families~\cite{Wang2015} in these simulation runs (colored lines with downwards pointing triangles as symbols), a family referring to all replicas descending from the same copy at $\beta_0$. For comparison, the (averaged) data from Fig.~\ref{fig:overview_-1}(a1) are shown as black dashed lines, as well as data from PA runs heating from $T=0$ (upward facing triangles).
Here, starting from $\beta\rightarrow\infty$ and annealing to $\beta\rightarrow 0$ is easily feasible, as for rectangular PBC the ground states can be sampled exactly: This is realized by randomly drawing a RW describing the ground state~\footnote{Here, we only sample the $2^{L}$ ground states in which the vertical axis is pinned, as this simplifies the implementation and already for moderate system sizes the other pinned directions are practically negligible.}.
When considering the specific heat [panel (a)], one sees the aforementioned sharp jumps for the individual runs. While these are absent in both the averaged cooling and heating runs, for inverse temperatures $\beta \gtrsim 1.3$ the data for heating and cooling clearly disagree and show a degree of hysteresis which is in agreement with the previously reported first-order-like character of these features.

Figure~\ref{fig:takeover_-1_L48}(b) shows the number of surviving families $f$ in these simulations. For the individual cooling runs that exhibited the spikes only very few families survived, with a strong decline in $f$ at the location of the spikes. In contrast, in the (averaged) cooling and heating runs without the spikes always several hundred families remained. While this observation usually suggests a good level of equilibration, the presence of hysteresis indicates that this is not the case for these simulation runs.
We conjecture that the mechanism for the hysteresis and the rare jumps may be the following: There is a small chance that replicas have no vertical stripes and thus have a much lower energy than the rest of the population. As this occurs only in some runs, the probability for this to happen is at most of the order of $1/R$. Thus, it is unlikely for more than a few replicas to find this lower energetic state, and the replicas take over the entire population. In the heating run, one starts without stripes percolating in the vertical direction while these tend to survive at lower temperatures upon cooling.

\begin{figure}[tb!]
  \includegraphics{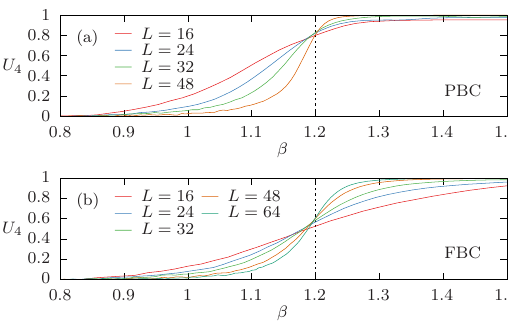}
  \includegraphics{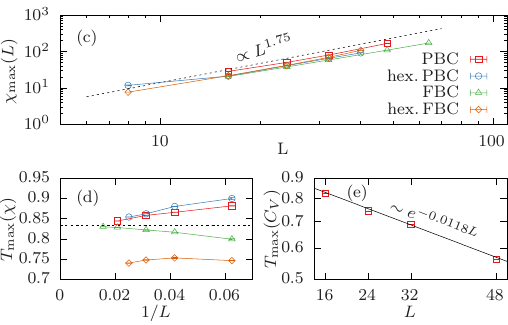}
  \caption{Finite-size scaling plots for $J_2=-1$. (a) and (b) show the nematic Binder parameter $U_4$ for rectangular PBC and FBC, respectively. (c) The susceptibility peak's height, (d) and its location $T_{\max}$. The horizontal dashed line in (d) corresponds to the potential transition temperature $T = 1 / 1.2 \approx 0.83$. Panel (e) shows the location of the lowest-temperature specific-heat peak, and the solid line indicates its consistence with an exponential decay towards zero as $L\rightarrow\infty$.\label{fig:fss_-1.0}}
\end{figure}

Although the accuracy of the data for the low-temperature features may be questionable, the potential transition around $\beta \approx 1.2$ shows no signs of being of first order, and in all runs had compatible outcomes.  The Binder parameter is consistent with a crossing at $\beta\approx 1.2$ for PBC and FBC, see Fig.~\ref{fig:fss_-1.0}. As the value of $U_4$ at the crossing is not a fully universal quantity, its value is expected to vary with boundary conditions~\cite{binder:81}.  Figures~\ref{fig:fss_-1.0}(c) and~\ref{fig:fss_-1.0}(d) show the finite-size scaling of the high-temperature peak of the susceptibility. As was the case for $J_2=-0.5$, its height is approximately compatible with a power-law growth with exponent $1.75$ for the different types of boundary conditions. Except for hexagonal FBC the locations of the maxima are consistent with approaching $T\approx 0.83$ as $L\rightarrow \infty$. Figure~\ref{fig:fss_-1.0}(e) shows that at least for the system sizes considered the location of the lowest-temperature specific-heat peak is consistent with an exponential decay with $L$ (solid line), although as discussed above at least for $L=48$ the level of equilibration is uncertain.

\subsection{Different aspect ratios}\label{sec:aspectRatio}

\begin{figure}[tb!]
  \includegraphics{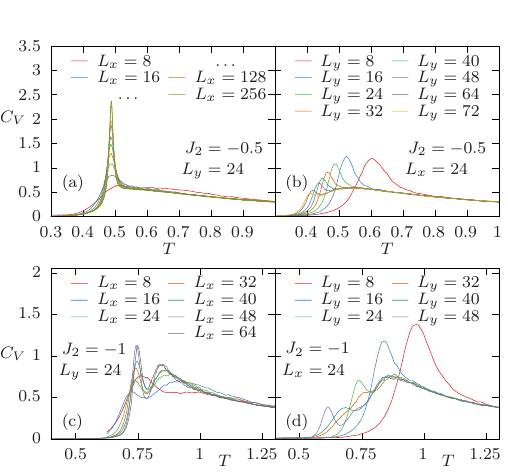}
  \caption{Specific-heat curves for different aspect ratios $L_x$:$L_y$. In panels~(a) and (c), $L_x$ is altered while keeping $L_y=24$ fixed and, vice versa, in panels~(b) and~(d) $L_y$ is varied at constant $L_x=24$. The system sizes omitted in the legend (indicated by $\dots$) of panel~(a) are $L_x=24, 32, 40, 48, 64,$ and $72$. In the top row [panels (a) and (b)] $\rv=-0.5$, and in the bottom row [panels (c) and (d)] $\rv=-1$. \label{fig:aspectRatio}}
\end{figure}

Typically, the aspect ratio is regarded as a minor detail in models with isotropic interactions strengths (as is the case here). Following this line of argument, little attention has been paid to the aspect ratio in previous studies of the $J_1$--$J_2$ Ising model. What is more, on the honeycomb lattice there is no canonical aspect ratio, and therefore various aspect ratios have been used --- and previous authors appear not to have been aware of such differences. In all cases, systems were specified by their linear system size $L$ (or their number of spins~$N$) and the chosen interaction strengths $J_1$ and $J_2$. Already a superficial comparison of data, e.g., for $J_2=-0.5$ from Ref.~\cite{Zukovic2022}, from Ref.~\cite{Azhari2025}, and from the present work shows that there are significant differences for data of the same system sizes~$L$. At first, one may be inclined to attribute this to the different respective Monte Carlo techniques. Instead, a direct comparison of the different simulation data shows that the vast majority of such differences are due to the use of different aspect ratios (see Appendix~\ref{app:directComparisonSimulations}).

In Table~\ref{tab:aspectRatioLiterature} we list all Monte Carlo studies known to us, in which this model is considered.  We define the aspect ratio with respect to the way we represent a honeycomb system as a two-dimensional array, that is an array of dimensions $L_x \times L_y$ representing a honeycomb system with $N=L_x L_y$ spins as shown in Fig.~\ref{fig:memory-vs-real-vs-brick}(a). Here, the $x$-direction is the armchair direction, and the $y$-direction the perpendicular zigzag direction.  While uniquely specifying the aspect ratio, a different array representation would result in different values; with other representations including the commonly used brick-wall lattice, viz.\ Fig.~\ref{fig:memory-vs-real-vs-brick}(c). Besides our aspect ratio denoted as $2L$:$L$, two more ratios $2L$:$L/2$ and $L$:$L$ have been used in the literature~\cite{Zukovic2020,Zukovic2021,Acevedo2021,Corte2021,Zukovic2022,Gessert2024,Gessert2025,Azhari2025,Li2025}, with the former corresponding to systems half as tall as ours (in the zigzag direction), and the latter to systems half as wide as ours (in the armchair direction).  Since most of the previous studies did not provide sufficient detail regarding the lattice shapes actually used in the simulations, these ratios were obtained by a combination of examining provided snapshots and comparing our simulation data directly against the corresponding published simulation results.  In fact, each ratio is ``square'' in its own way. Our aspect ratio $2L\times L$ corresponds to a system of $L\times L$ hexagons, and the aspect ratio of the displayed snapshots is closest to 1. The ratio $2L$:$L/2$ corresponds to a ratio of $L$:$L$ in the brick-wall representation. And last, $L$:$L$ is square in our memory coordinates.

\begin{table}[b]
     \caption{Different aspect ratios $L_x:L_y$ used in the literature, as well as simulation methods and the studied ranges in $J_2$. $L_x$ refers to the length in the armchair direction, and $L_y$ to the length in the orthogonal zigzag direction.\label{tab:aspectRatioLiterature}}
    \begin{ruledtabular}
    \begin{tabular}{rrrcccc}
        \multicolumn{1}{c}{\multirow{ 2}{*}{Ref.}}&\multicolumn{1}{c}{\multirow{ 2}{*}{Method}}& \multicolumn{3}{c}{$J_2$ vs.\ $-\frac{1}{4}$} & \multicolumn{1}{c}{\multirow{ 2}{*}{$L_x\,:$}} & \multicolumn{1}{c}{\multirow{ 2}{*}{$L_y$}}\\
        &  & \multicolumn{1}{c}{$<$} & \multicolumn{1}{c}{$=$} &\multicolumn{1}{c}{$>$}&&\\\colrule
        \multicolumn{1}{c}{this work} & PA&$\times$& $\times$ &&$2L$&$L$\\
        Li\ \emph{et al.} (2025)~\cite{Li2025} & Metropolis & $\times$	& $\times$ &$\times$&$2L$&$L/2$\\
        Azhari\ \emph{et al.} (2025)~\cite{Azhari2025} & W.-L. & $\times$	& $\times$ &$\times$&$2L$&$L/2$\\
        Gessert\ \emph{et al.} (2025)~\cite{Gessert2025} & PA ($n$-fold) & 	&  &$\times$&$2L$&$L$\\
        Gessert\ \emph{et al.} (2024)~\cite{Gessert2024} & PA& 	&  &$\times$&$2L$&$L$\\
        {\v{Z}}ukovi{\v{c}} (2022)~\cite{Zukovic2022} & PT &$\times$&&&$L$&$L$\\
        {\v{Z}}ukovi{\v{c}} (2021)~\cite{Zukovic2021} & PT & &&$\times$&$L$&$L$\\
        Acevedo\ \emph{et al.} (2021)~\cite{Acevedo2021} & Metropolis &$\times$&&$\times$&$2L$&$L/2$\\
        Corte\ \emph{et al.} (2021)~\cite{Corte2021} & Metropolis & &$\times$&$\times$&$2L$&$L/2$\\
        {\v{Z}}ukovi{\v{c}}\ \emph{et al.} (2020)~\cite{Zukovic2020} & Metropolis &$\times$&&&$L$&$L$\\
    \end{tabular}
    \end{ruledtabular}
\end{table}

We now consider the effects of such different aspect ratios on the thermodynamic features of the system. In Fig.~\ref{fig:aspectRatio} we show the specific heat $C_V$ when fixing the length in one direction, and varying it in the other. The most immediate observation is that the effect of changing $L_x$ or changing $L_y$ (while keeping the other fixed) is rather different despite couplings not explicitly depending on the direction. Recall that most ground states are described by RWs spanning the system horizontally, thus clearly being anisotropic in nature. More specifically, $L_x$ sets the length of the individual RWs, and $L_y$ the number of parallel RWs. Due to the anisotropy of the ground states, it is plausible that also the behavior for $T>0$ is susceptible to different aspect ratios.

For $J_2=-0.5$, when fixing $L_y$ [panel~(a)] the peak locations change only moderately whereas the heights increase.  It is worth pointing out that for finite $L_y$ and $L_x\rightarrow \infty$ the system becomes effectively one-dimensional, and no phase transition is expected when $L_x \rightarrow \infty$, even though the data appear to be suggestive of a transition.  At constant $L_x$ [panel~(b)], on the other hand, the peak in $C_V$ moves towards lower temperatures and becomes smaller with increasing $L_y$.

Similarly, for $J_1=-1$ [panels~(c) and~(d)], increasing $L_x$ leads to sharper peaks in the specific heat, and increasing $L_y$ shifts features towards lower temperatures. What is different from the previous scenario is that also the number of peaks increases with $L_y$.  In this case, the peak heights vary non-monotonously with $L_y$.

\section{Discussion}\label{sec:discussion}

So far we held back with discussing our numerical findings regarding $\rv<-1/4$ as their interpretation is not straightforward. The discussion in the present section will focus on the signs of a phase transition at $\beta \approx 2.46$ and $\beta \approx 1.2$ for $J_2 =-0.5$ and $J_2=- 1$, respectively.  The additional features suggestive of transition behavior seen for $J_2=- 1$ at lower temperatures move very rapidly towards lower temperature, and it appears likely that these points reach to $T=0$ as $L\rightarrow\infty$. Their locations appeared to be consistent with an exponential decay towards $T=0$ for increasing system size $L$. However, in this regime we saw clear indications that our simulations failed to equilibrate properly at least for $L=48$. Therefore, without further investigations employing techniques more suitable for first-order phase transitions, other scenarios for the low-temperature thermal behavior cannot be safely ruled out.

For the potential transition at temperatures $\beta \approx 2.46$ ($J_2 =-0.5$) and $\beta \approx 1.2$ ($J_2=- 1$), we find that there are three possible scenarios: i) a crossover without any singularities, ii) a continuous phase transition, or iii) a BKT transition.  Unfortunately, given the strong scaling corrections, the sensitive dependence on boundary conditions and aspect ratios, and the slow dynamics connected to the inherent frustration, our data do not allow us to reach a final decision for one of the above scenarios. Therefore, we will present arguments for each of the three scenarios, and in each case we highlight similarities to other models.

\emph{i) Crossover.}  Particularly for $J_2 = -0.5$ when only considering PBC, the sharp specific-heat peaks were previously seen as a strong indicator for a phase transition~\cite{Zukovic2020,Acevedo2021,Zukovic2022,Azhari2025}. For FBC, however, the feature is entirely absent, and even for PBC the specific-heat peak starts to decrease in size beyond a certain system size. Interestingly, very similar behavior is seen for the two-dimensional gonihedric model (see Appendix~\ref{app:gonihedricModel}) which allows for an analytical solution of the plaquettes-only case ($\kappa = 0$): Also there a sharp peak is visible for PBC and absent when using free FBC.  From the analytical solution in this case it is clear that there are no singularities as $L\rightarrow \infty$ and hence no thermodynamic phase transition.  On the one hand, this picture is obfuscated by the fact that unlike in the gonihedric model, $C_V$ for FBC in the present model is \emph{not} a trivial function of temperature. On the other hand, this non-trivial behavior for FBC may be expected as the same is already observed for the Ising AFM on the triangular lattice (corresponding to $J_1=0$ here)~\cite{Millane2006,Kim2015}, for which it is known that there is no phase transition~\cite{Wannier1950,*Wannier1973Errata}.

Closely related to the gonihedric model is the frustrated $J_1$-$J_2$ Ising model on the square lattice for $J_2=-|J_1|/2$, which is equivalent to the Hamiltonian (\ref{eq:gonihedricModel}) for $\kappa = 1$. Reference~\cite{Kalz2008} finds a similar structure of the specific heat (using periodic boundary conditions), concluding that the transition is suppressed to $T=0$. This conclusion is compatible with Ref.~\cite{Espriu2004}, where it is argued that the gonihedric model is thermodynamically trivial not just for $\kappa = 0$ but also for any $\kappa \neq 0$.
In addition, the ground states for the $J_1$-$J_2$ Ising model on the square lattice for $J_2=-J_1/2$ and ferromagnetic $J_1>0$ in fact are very similar to the ground states seen here: In the same way as lines of pairs of spins along one of the three principal axes can be flipped at no cost after pinning one of the axes, here, in the model on the square lattice lines of single spins along either the horizontal or the vertical direction can be flipped at no energetic cost (starting from the ferromagnetic state), giving rise to $2\times 2^L$ states. As is the case here, in doing so one counts states twice, namely the two ferromagnetic states, yielding a degeneracy of $2^{L+1}-2$ for an $L\times L$ system~\cite{Kalz2008}.
Further, a similar order parameter to the nematic one in Eq.~\eqref{eq:nematicOP}, e.g., as the one used in Refs.~\cite{Booth1995,Ye2021}, can distinguish between the ground state and high-temperature disordered state, despite most evidence indicating the absence of a phase transition. Although it is not clear whether the temperature behavior would be similar to the one observed here for $\langle|\eta|\rangle$, it highlights that a plausible order parameter candidate alone does not suffice to conclude the existence of a phase transition.

Another indication for a crossover scenario is the similarity to a special case of the anisotropic Ising AFM on the triangular lattice, namely when two antiferromagnetic couplings are equal and the third is stronger, which results in the same kind of ground states as the ones seen here~\cite{Dublenych2013}. While Houtappel~\cite{Houtappel1950} solved the anisotropic Ising AFM in general, this particular case was looked at in more detail only somewhat recently~\cite{Hotta2011}. Similar to the case here, the specific heat shows nontrivial behavior, but from the analytical solution it is clear that asymptotically there are no singularities. 
Since for $J_1=0$, the present model acts as two uncoupled (isotropic) Ising AFM on the triangular lattice, it appears plausible that setting $J_1\neq 0$ could result in an Ising AFM with effectively anisotropic couplings.
A related model, albeit not exactly solved, is the Ising AFM on an elastic triangular lattice, which has been discussed in the context of buckled colloidal monolayers~\cite{Han2008,Shokef2009,Shokef2011,Zhou2017} and for which Ref.~\cite{Shokef2011} reported on glassy dynamics reminiscent of what has been reported for the present model~\cite{Zukovic2020}; see Ref.~\cite{Dublenych2013} arguing that both models may provide an effective description of buckled colloidal monolayers.

Finally, the absence of a clear crossing of the Binder parameter particularly for PBC and $J_2 = -0.5$ [cf. Fig.~\ref{fig:fss_-0.5}(a)] is a further indication for the absence of a phase transition. In fact, $U_4$ seen for PBC and using $J_2 = -0.5$ has a certain similarity with the Binder parameter of the one-dimensional Ising model, see Appendix~\ref{app:binderParam1DIsing}. %

\emph{ii) Continuous phase transition of finite order.}  While the mere existence of a plausible order parameter clearly does not prove the existence of a phase transition, numerically $\langle|\eta|\rangle$ shows a clear signal indicative of a transition. Also, the finite-size scaling analysis of the related nematic susceptibility is consistent with a scaling of its maximum $\sim L^{1.75}$. While at first sight the observed value of $\nu > 1$ might be taken as an indicator for a crossover, it merely implies a negative $\alpha$ through the hyperscaling relation $\nu d = 2-\alpha$.  This is consistent with the behavior of the specific-heat peaks, which for both $J_2 = -0.5$ and $J_2 = -1$ appear to be non-divergent.
Assuming that the secondary peak in $C_V$ for FBC seen for $J_2=-0.5$ that only appears for the largest system sizes is linked to asymptotic criticality, it also appears plausible that $U_4$ is subject to strong scaling corrections resulting in poor crossing behavior for the considered system sizes.

\emph{iii) BKT-type transition.} 
The main reason to suspect a BKT transition is the fact that the encountered spin configurations show a surprising similarity to the ones seen in the dipolar Ising model~\cite{Vedmedenko1998,Simmons2024}, for which there is analytical~\cite{Abanov1995} and numerical~\cite{Bab2019} evidence for the occurrence of BKT transitions. 
What is more, the connection of the present model to the Ising AFM on the triangular lattice could also hint at the existence of a BKT transition: As is well known, the (isotropic) Ising AFM on the triangular lattice remains disordered at all temperatures, with a critical point at $T=0$ at which the correlation functions decay algebraically~\cite{Stephenson1964,Stephenson1970}. 
In some cases, introducing further terms to the spin Hamiltonian gives rise to a critical phase separated by a BKT-like transition such as when adding an external-magnetic-field term~\cite{Bloete1991,Bloete1993,Qian2004} or ferromagnetic NNN interactions~\cite{Nienhuis1984, Landau1983,Kitatani1988,Qian2004}, but see Ref.~\cite{Korshunov2005} finding signs of a first-order phase transition when both NN and NNN interactions are antiferromagnetic.
Since in the absence of NN interactions ($J_1=0$) the present model corresponds to the Ising AFM on the triangular lattice, it appears plausible that $J_1 \neq 0$ could result in a critical phase separated by BKT transitions. Note that the apparent contradiction with the analogy to the anisotropic AFM not showing a phase transition is resolved by the fact, that also that system is said to exhibit a BKT transition when ferromagnetic NNN interactions are added~\cite{Sato2013}.

At this stage, however, it is impossible to demonstrate the existence of a BKT transition with certainty. (Even seeing the true BKT character for the \emph{XY} model numerically has proven a rather challenging task~\cite{Janke1991,Janke1993a,Hasenbusch2008,Hasenbusch2009}.)
Nonetheless, there are numerous observations consistent with a BKT-type transition. First of all, topological defects are clearly seen (although their effective interaction remains unclear). Further, the scaling of the nematic susceptibility is consistent with the expected $L^{2-\eta}$ behavior where $\eta=1/4$. The expected weak system-size dependence of the specific heat is seen at least for free boundary conditions, and also the non-crossing behavior of $U_4$ shifting towards lower temperatures as seen when employing periodic boundary conditions is consistent with behavior seen in the \emph{XY} model.

\section{Summary and outlook} \label{sec:conclusion}
We have studied the $J_1$-$J_2$ Ising model on the honeycomb lattice for $J_2 = -1/4$, $J_2 = -0.5$ and $J_2 = -1$, for which the ground state is largely degenerate. For the highly symmetrical point $J_2 = -1/4$, where ferromagnetic and striped configurations become degenerate, we find a thermodynamically trivial system without a phase transition. The degeneracy leads to a multitude of ground states with a non-zero ground-state entropy per site, for which we provide an accurate estimate $S(T=0)/N = \resultZeroTEntropyMinusQuarter$ using a tailored population annealing technique.  For $J_2 < -1/4$, the degeneracy was calculated analytically for periodic boundary conditions both in the rectangular and hexagonal embeddings. In both cases the ground-state entropy scales linearly with the system size $L$.  It was demonstrated that in the ground state the threefold lattice symmetry is broken, thus raising the question of a potential finite-temperature phase transition. A nematic order parameter $\eta$ linked to this symmetry was proposed.

Moving away from ground states to non-zero temperatures, we used population annealing Monte Carlo simulations to investigate the temperature behavior and the potential phase transition. As the relevant spin symmetry is linked to that of the lattice, we considered various types of boundary conditions to see whether some of the thermodynamic signals might be induced by the boundaries. We indeed find that the previously reported sharp peaks~\cite{Zukovic2020,Zukovic2022} in the specific heat are strongly suppressed when using free instead of periodic boundary conditions. For both coupling ratios $J_2 = -0.5$ and $J_2 = -1$ considered here, there exists a temperature for which both the specific heat and the nematic order parameter show a clear signal throughout the different types of boundary conditions. The relevant specific-heat peak for $J_2=-0.5$ is only visible for the largest system sizes and therefore was not seen in previous work~\cite{Zukovic2020,Azhari2025}, whereas for $J_2=-1$ it is seen already for smaller system sizes and agrees with that reported in Ref.~\cite{Zukovic2022}.

This potential transition at $T\approx0.407$~$(\beta \approx 2.46)$ and $T\approx 0.83$~$(\beta \approx 1.2)$ for $J_2=-0.5$ and $J_2=-1$, respectively, was studied using a finite-size scaling analysis. In both cases the nematic susceptibility was consistent with a power-law growth $\chi\sim L^{1.75}$. Based on the numerical data it was not possible to unambiguously discern between genuine critical behavior and a cross-over without true thermodynamic singularities. If one assumes a continuous phase transition, then both the location of the nematic susceptibility and the scaling of the height of the specific heat are consistent with a negative $\alpha$-exponent.
What is more, the model shows surprisingly similar behavior to the dipolar Ising model in some cases~\cite{Vedmedenko1998,Simmons2024}, which is considered to have two BKT transitions.
Thus, a BKT transition cannot be ruled out, and in fact the scaling of $\chi$ is consistent with the expected value $\eta=1/4$ of BKT transitions.

Besides this leading transition-like behavior, numerous features in the specific heat and the nematic order parameter were observed, which depend much more strongly on the chosen boundary condition, particularly for $J_2=-1$. Although consistent with these features approaching zero temperature in the thermodynamic limit, our simulations showed clear signs of failing to equilibrate at these lower temperatures. This is caused by insufficient histogram overlap of neighboring temperatures due to the bimodal distribution of the energy histograms. This was also reported in Ref.~\cite{Zukovic2022}, describing these features as first-order-like. To better understand this effect, other simulational setups such as multicanonical methods~\cite{Berg1991,Berg1992} or microcanonical population annealing~\cite{Rose2019,Pfaff2026} could be usefully employed.

One interesting aspect for future work would be studying the effect of introducing third-nearest neighbor interactions $J_3$.
Introducing the $J_3$ coupling removes the large degeneracy of the low-temperature phase~\cite{Dias2023}, and instead it is found that for the relevant range of $J_2$, the model exhibits a striped ground state with orientational order corresponding to a value of two for the absolute value of the nematic order parameter $\eta$ instead of one. It would be interesting to see whether the partially-ordered low-temperature phase of meandering stripes observed here with $J_3=0$ is visible for $J_3>0$ as an intermediate phase. A similar behavior is known, e.g., for the dipolar Ising model~\cite{Booth1995}, in which the tetragonal or nematic phases at high temperatures transit towards the low-temperature stripe phase.

\begin{acknowledgments}
  D.G. and W.J. were supported by the Deutsch-Franz\"osische Hochschule (DFH-UFA) through the Doctoral College ``$\mathbb{L}^4$'' under Grant No.\ CDFA-02-07. They further acknowledge support by the Leipzig Graduate School of Natural Sciences ``BuildMoNa''. D.G. and M.W. thank the DAAD (Project ID 57807776) and SPARC (SPARC-GIANT/2025-2026/PG250011) for support through an Indo-German grant scheme.
\end{acknowledgments}

\appendix

\section{Minimization of surface energy for free boundary conditions}\label{app:energyMinFBC}

\begin{figure}[tb!]
  \includegraphics{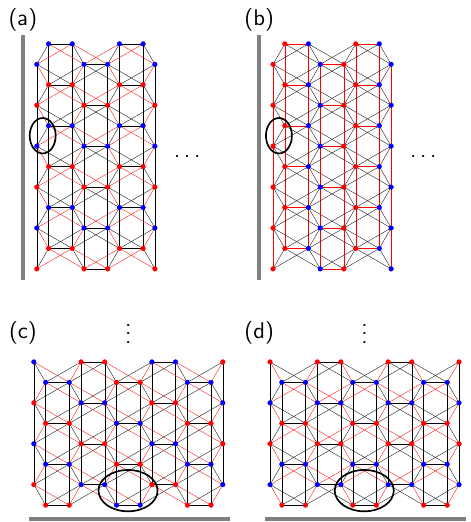}
  \caption{Schematic low-energy surface configurations for zigzag [(a) and (b)] and armchair boundaries [(c) and (d)]. (a) and (b) show antiferromagnetic and ferromagnetic zigzag boundaries, respectively.  (c) and (d) show outer spins of alternating sign, and with the same sign, respectively. Red and blue circles denote Ising spins $-1$ and $1$. Frustrated interactions are denoted by solid red lines. The ellipses mark the primary cell used for energy calculation. States (b) and (d) are preferred for $J_2/J_1\in(-1/4,-1/2)$, and (a) and (c) are preferred for $J_2/J_1 < -1/2$.\label{fig:boundaryConfigs}}
\end{figure}

In the main text we observed that for free boundaries the ground states for $-\infty < \rv < -1/4$ further depend on $\rv$. Although precisely determining the set of ground states appears unfeasible due to its non-trivial system-size dependence, it is possible, however, to locally minimize the surface energy. (The non-trivial system-size dependent task then becomes how to combine surface configurations with bulk configurations, which already for the Ising AFM on the triangular lattice can only be done for small lattices~\cite{Millane2006,Kim2015}.) As became apparent from considering the configurations in Fig.~\ref{fig:overviewFbcSnaps}, the configurations of the surface spins minimizing the energy depend on $J_2$ (in addition to the already existing condition that $J_2 < -1/4$).

Figure~\ref{fig:boundaryConfigs} shows the found low-energy surface configurations for both zigzag and armchair boundaries. Which configuration is energetically more favorable can easily be determined by counting the number of broken $J_1$ and $J_2$ bonds (solid red lines). A broken $J_1$ ($J_2$) bond is associated with an energy cost of $2J_1$ ($-2J_2$).  We use a general $J_1 > 0$ (even though we set $J_1=1$) to better distinguish energetic cost associated to {\nn} and {\nnn} interactions.  It suffices to count bonds in the primitive cell highlighted by ellipses with interactions crossing the ellipses only counting as a half interaction. In (a), one counts one broken $J_1$ interaction and one broken $J_2$ interaction, whereas in (b), there are $1/2$ broken $J_1$ bonds and two broken $J_2$ bonds. Thus, the energy difference of the two states is \begin{equation} e_{(a)} - e_{(b)} \sim (J_1 - J_2) - (J_1/2 - 2 J_2) = J_1/2 + J_2, \end{equation} which is negative when $J_2<-J_1/2$. Hence, zigzag boundaries tend to order ferromagnetically for $J_2/J_1\in (-1/4,-1/2)$, and antiferromagnetically for $J_2 < -J_1/2$.

Similarly, for armchair boundaries one counts $1.5$ broken $J_1$ bonds and $3$ broken $J_2$ bonds in (c), and two broken $J_1$ bonds and two broken $J_2$ bonds (d), yielding
\begin{equation}
  e_{(c)} - e_{(d)} \sim (3/2 J_1 - 3 J_2) - (2 J_1 - 2 J_2) = -J_1/2 - J_2.
\end{equation}
This expression is negative whenever $J_2 > -J_1/2$. Thus, pairs of surface spins with alternating sign (c) are preferred for $J_2<-J_1/2$, while for $J_2/J_1\in (-1/4,-1/2)$ surface spins of the same sign are energetically more favorable.

\section{Population annealing estimates for the zero-temperature entropy}\label{app:groundStateEntropy}

\begin{table}[b]
  \caption{Estimates for the zero-temperature entropy per~$L$ for the system with rectangular PBC. Results from PA runs with population size $R$ were averaged over $M$ runs; for comparison we also show the values of Eq.~\eqref{eq:groundStateDegPeriodicRect} and the estimate that results when only considering the leading contribution from states with the vertical axis pinned and the two vertical states, as was done in Ref.~\cite{Azhari2025}.\label{tab:pa_gs_entropy_rect}}
  \begin{ruledtabular}
    \begin{tabular}{rrrlrr}
        \multicolumn{1}{c}{$L$}	& \multicolumn{1}{c}{Eq.~\eqref{eq:groundStateDegPeriodicRect}} & \multicolumn{1}{c}{Ref.~\cite{Azhari2025}} & \multicolumn{1}{c}{PA} &	\multicolumn{1}{c}{$R$} & \multicolumn{1}{c}{$M$} \\\colrule
        $4$	& $0.72259\dots$ &	$0.72259\dots$	& &&\\
        $6$	& $0.72612\dots$ && $0.72603(7)$ & $10^6$ & $10$ \\
        $8$	& $0.70524\dots$ &	$0.69412\dots$	& $0.70541(13)$ & $10^6$ & $10$ \\
        $10$ & $0.69903\dots$ && $0.69894(7)$ & $10^6$ & $10$ \\
        $12$	& $0.69559\dots$ &$0.69319\dots$& $0.69560(6)$ & $10^6$ & $10$ \\
        $16$	& $0.69363\dots$ &	$0.69315\dots$	& $0.69356(5)$  & $2\times10^6$ & $10$ \\
        $24$	& $0.69317\dots$ &	$0.69315\dots$	& $0.69317(5)$  & $2\times 10^6$  & $15$ \\
        $\infty$ & \multicolumn{2}{c}{$\ln(2)\approx 0.69314718\dots$} &&&			
    \end{tabular}
  \end{ruledtabular}
\end{table}

To numerically test the validity of Eqs.~\eqref{eq:groundStateDegPeriodicRect} and \eqref{eq:groundStateDegPeriodicHex} we have carried out PA simulations of small system sizes with very large populations to obtain the ground-state entropy. Here, the parameters $\alpha^*$ for the adaptive temperature schedule and $\rho^*$ for the adaptive sweep schedule were chosen larger than in the rest of the simulations to achieve maximal accuracy, that is $\alpha^*= 0.87 - 0.9$, and $\rho^*=0.95$. All simulations were carried out at $\rv=-0.5$. The procedure in obtaining $S(T=0)$ is the same as in Sec.~\ref{sec:minusOneQuarter} using Eq.~\eqref{eq:pa_S_GS_estimator}.

The results concerning rectangular PBC are summarized in Table~\ref{tab:pa_gs_entropy_rect}. In Ref.~\cite{Azhari2025} the $2^L$ ground states with horizontal stripes and the two configurations with straight vertical stripes were described, but the remaining states in which axis 2 or 3 is pinned were neglected. 
Thus, one arrives at $2^L+2$ for the number of ground states, which is the expression used in Table~\ref{tab:pa_gs_entropy_rect} for Ref.~\cite{Azhari2025}, and which clearly differs from Eq.~\eqref{eq:groundStateDegPeriodicRect}. We have enumerated all $2^{32}$ states exactly for $L=4$, and have counted $18$ ground states. In this case both expressions happen to coincide with this result, i.e., $2^L+2=2^L+2\times 2^{L/2}-6=18$.  In contrast, for $L\neq 4$ the two forms differ, while approaching the same limit for $S(T=0)/L$ as $L\rightarrow\infty$. For $6\leq L \leq 16$ the values for $S(T=0)/L$ obtained using PA in both cases are just outside the given one-$\sigma$ error bars of the values from Eq.~\eqref{eq:groundStateDegPeriodicRect}, and clearly different from the value $2^L+2$ of Ref.~\cite{Azhari2025}. Since Ref.~\cite{Azhari2025} only considers system sizes for which $L$ is divisible by 4, we have left the entry blank when $L$ is not divisible by 4. In fact, for $L=6$ and $L=8$ the accuracy is good enough to numerically determine the exact number of ground states. When carrying out this calculation, one obtains 77.96(4) and 282.4(3) for the number of ground states from our PA simulations for $L=6$ and $L=8$, which, when imposing that this number has to be an even integer leads to the exact values of Eq.~\eqref{eq:groundStateDegPeriodicRect}, i.e., $88$ and $282$.
Note that this also shows numerically that the states counted twice were taken into account correctly by $n_\text{oc}$, which is different depending on whether $L$ is divisible by 4 or not.
Further, we have carried out an analogous analysis for the system with hexagonal PBC, cf.\ Table~\ref{tab:pa_gs_entropy_hex}. Also there, our estimates from PA are in excellent agreement with the values calculated via Eq.~\eqref{eq:groundStateDegPeriodicHex}.

\begin{table}[b]
  \caption{Estimates for the zero-temperature entropy per $L$ for the system with hexagonal PBC. Results from PA runs with population size $R$ averaged over $M$ runs are compared to the values of Eq.~\eqref{eq:groundStateDegPeriodicHex}.\label{tab:pa_gs_entropy_hex}}
    \begin{ruledtabular}
    \begin{tabular}{rrlrr}
        \multicolumn{1}{c}{$L$}	& \multicolumn{1}{c}{Eq.~\eqref{eq:groundStateDegPeriodicHex}} & \multicolumn{1}{c}{PA} &	\multicolumn{1}{c}{$R$} & \multicolumn{1}{c}{$M$} \\\colrule
        $8$	& $0.46721\dots$ & $0.46712(9)$ & $10^6$ & $10$ \\
        $16$	& $0.41475\dots$ 	& $0.41473(5)$ & $10^6$ & $20$ \\
        $24$	& $0.39233\dots$   & $0.39237(5)$ & $10^6$  & $15$ \\
        $\infty$ & $\ln(2)/2\approx 0.34657\dots$ &&&			
    \end{tabular}
    \end{ruledtabular}
\end{table}

\section{Zero-temperature value of the nematic order parameter}\label{app:etaZeroTemp}

Assuming stripes are parallel with respect to one of the three directions, and otherwise randomly follow one or the other direction, the low-temperature value of 1 for the nematic order parameter $\eta$ can be understood as follows. Without loss of generality, we can assume that stripes are parallel with respect to the $(0,1)$ axis, that is the $k=1$ contribution in the sum~(\ref{eq:nematicOP}) is always $-1$.  Next, we note that $\sigma_j \sigma_{j\oplus k}$ for $k=2$ and $k=3$ are always opposite in sign, and so are $\sigma_j \sigma_{j\ominus k}$. Thus, for low-temperature stripe states with stripes parallel to $(0,1)$, $\eta_j = -1 + (\sigma_j \sigma_{j\oplus 2} + \sigma_j \sigma_{j\ominus 2}) \frac{\sqrt{3}}{2}i$. The expression $(\sigma_j \sigma_{j\oplus 2} + \sigma_j \sigma_{j\ominus 2})$ itself takes values of $-2$, $0$, and $2$ at random with $-2$ and $2$ having the same probability. Hence, in the thermodynamic limit it averages to zero, and $\langle |\eta| \rangle$ (asymptotically) is equal to $1$.

Continuing the same random walk argument for {\em finite\/} (periodic) systems, this term does not actually average away but instead adds a contribution $\sim 1 / \sqrt{L}$ to the imaginary part of $\eta$, and hence a $\sim 1/L$ contribution to $\langle |\eta|\rangle$. Specifically, for rectangular PBC of an $L \times L$ system, the ground state can be understood as a one-dimensional random walk of length $L-1$, in which $L$ parallel horizontal stripes can at each column either turn up or down.  Using $(j\oplus k) \ominus k = j$, we can simplify $|\eta|$ as
\begin{equation}
  |\eta| = \left| \frac 1 N \sum_j \eta_j \right|= \left|\frac 1 N \sum_j \sum_{k=1}^3 \sigma_j \sigma_{j\oplus k} f_k\right|.
\end{equation}
In this setting, with the $k=1$ axis taking the role of the frozen-out axis, this becomes
\begin{equation}
   |\eta| = \left| \frac 1 N \sum_j (-1 + \sigma_j \sigma_{j\oplus 2} \sqrt{3} i) \right|.
\end{equation}
As the state is encoded by a random walk of length $L-1$ with $2L$ parallel ``copies'' the absolute value of the complex part on average is $\sqrt{3 (L-1)} \times 2L$, where it was used that the expected end-to-end distance of a random walk of length $L-1$ is $\sqrt{L-1}$. From this, the expectation value $\langle |\eta|\rangle$ at zero temperature is obtained as
\begin{equation}
  \begin{split}
    \langle | \eta |\rangle &= \frac{\sqrt{4 L^4 + 12 (L-1) L^2}}{N} = \sqrt{1+3L^{-1}-3L^{-2}}\\
    & = 1 + \frac 3 2 L^{-1} - \frac{21}{8} L^{-2} + O(L^{-3})\label{eq:etaZeroTempRectPBC}
  \end{split}
\end{equation}
using $N = 2L^2$. This relation was also numerically verified (not shown), and explains the values larger than $1$ observed for $\langle |\eta| \rangle$ that we report in Fig.~\ref{fig:overview_-0.5}(a2) of the main text.

\section{Different representations of the honeycomb lattice}\label{app:representationHoneycomb}
\begin{figure}
  \includegraphics{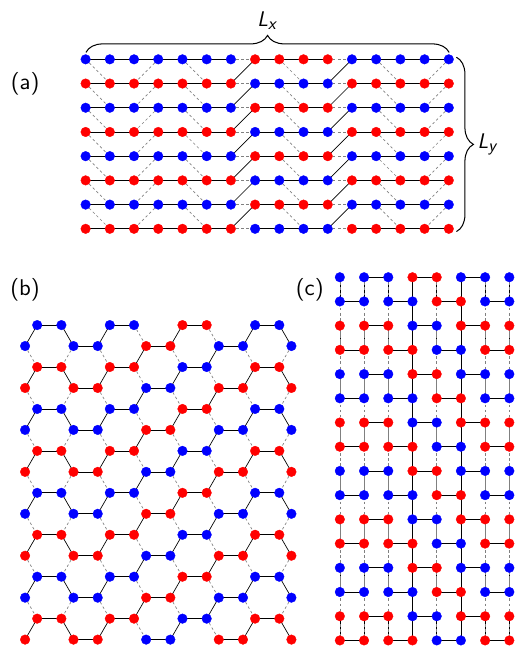}
  \caption{An $L=8$ honeycomb lattice (a) in the memory representation, (b) in physical coordinates, and (c) in the brick-wall representation. In each case, the same spin configuration is shown, which is a randomly chosen ground state for $\rv<-1/4$ of the system with PBC. Note that the brick-wall representation shown here is rotated by 90\textdegree{} compared to its usual definition. $L_x$ and $L_y$ used to define the aspect ratio are illustrated in panel (a).\label{fig:memory-vs-real-vs-brick}}
\end{figure}

Figure~\ref{fig:memory-vs-real-vs-brick} shows a striped ground state for $\rv < -1/4$ for an $L=8$ system with PBC in different lattice representations: (a) as addressed in memory in our simulations, (b) in the canonical honeycomb lattice representation, and (c) in the brick-wall representation. Note that in its usual form the brick-wall lattice is rotated by 90\textdegree.

The aspect ratio discussed here always refers to the ratio $L_x : L_y$ from the memory coordinates (which in our case, unless stated otherwise, is always 2:1).

\section{Percolation properties for $J_2=-0.5$}\label{app:percolation}
We consider the percolation properties for $J_2=-0.5$, and using (rectangular) PBC. The simulations were carried out using population annealing with a rejection-free update~\cite{Bortz1975} instead of the GPU update of the main text, see Ref.~\cite{Gessert2025} where this simulation method is used and discussed. Clusters are constructed from satisfied nearest-neighbor interactions, and are identified using the Hoshen-Kopelman algorithm. A configuration is identified as percolating in the horizontal (vertical) direction if there exists a cluster that spans the system in the respective direction. At each temperature this check is carried out and the wrapping probabilities $r_x$ and $r_y$ are determined through the population average of ones ($\exists$ a wrapping cluster) and zeros ($\nexists$ a wrapping cluster).

\begin{figure}[tb!]
  \centering
  \includegraphics{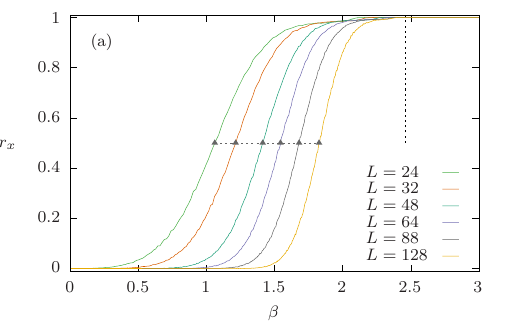}
  \includegraphics{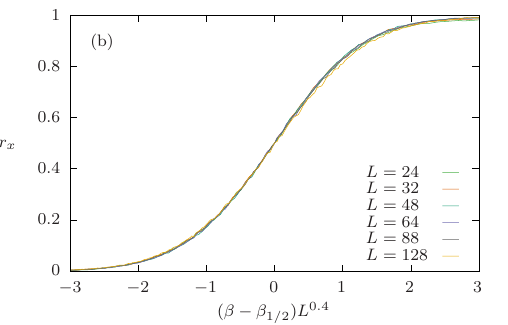}
  \caption{(a) Horizontal wrapping probability $r_x$ for different system sizes $L$. The observed intercepts $r_x \stackrel{!}{=} 0.5$ are indicated by triangles at $(\beta_{1/2},1/2)$ and connected with dashed lines. The vertical dashed line corresponds to the potential inverse transition temperature $\beta = 2.46$ of the main text. (b) Data of (a) collapses when plotted vs.\ $(\beta-\beta_{1/2})L^{0.4}$.\label{fig:wrapProb_x_J2-0.5}}
\end{figure}

Figure~\ref{fig:wrapProb_x_J2-0.5}(a) shows the measured wrapping probability in horizontal direction as a function of inverse temperature $\beta$ for various system sizes $L$. It can be seen that for all sizes configurations do not percolate at high and do percolate at low temperature. The temperature at which the system changes from non-percolating to percolating increases with system size. Where the wrapping probability crosses $1/2$ for each $L$ is marked by triangles. The corresponding $\beta$-value is denoted by $\beta_{1/2}$.

For a more quantitative picture the different curves were shifted by $\beta_{1/2}$ and rescaled by $L^{1/\tilde{\nu}}$ ($\tilde{\nu}$ being unknown here~\footnote{Note that $\tilde{\nu}$ is not expected to be identical to $\nu$ as we consider geometrical clusters. For $J_2=0$, $\nu=1$ and $\tilde{\nu}=15/8$ is expected~\cite{Stella1989,Janke2004}}). Numerically, we find the best collapse for $1/\tilde{\nu}\simeq0.4$, as shown in Fig.~\ref{fig:wrapProb_x_J2-0.5}(b).  Note that the exponent $\tilde{\nu}$ is related to the scaling exponent of the size of the largest cluster $\sigma$ by $\sigma d_f = 1/\tilde{\nu}$, where $d_f$ is the fractal dimension. As clusters here are stripes of width one, we expect $d_f \simeq 1$, and therefore simply $\sigma = 1/\tilde{\nu}$. Surprisingly, a value of $0.4$ for $\sigma$ is rather close to the $\sigma$-exponent of two-dimensional standard percolation, i.e., $\sigma = 36/91 \approx 0.396$ (although it is unclear why this exponent should apply here).

\begin{figure}[tb!]
  \centering
  \includegraphics{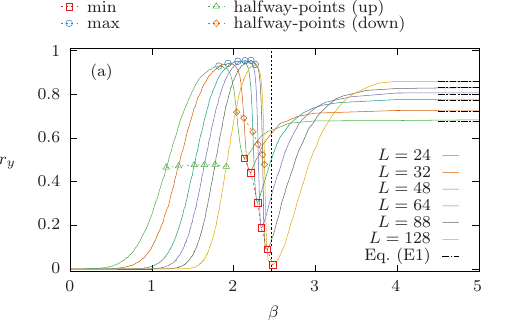}
  \includegraphics{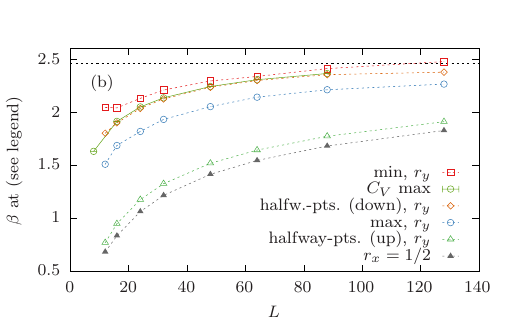}
  \caption{(a) Vertical wrapping probability $r_y$ for different system sizes $L$. Special points are highlighted and their labels are above the figure. The horizontal dashed-dotted lines for large $\beta$ show the $\beta\rightarrow\infty$ limit~\eqref{eq:zeroTempPercProbY}. (b) Special $\beta$ points from (a), as well as for $r_x$ of Fig.~\ref{fig:wrapProb_x_J2-0.5}(a), and the location of the specific heat maxima from Fig.~\ref{fig:overview_-0.5}(a1) of the main text. The dashed line [vertical in (a) and horizontal in (b)] corresponds to the potential inverse transition temperature $\beta = 2.46$ of the main text.\label{fig:wrapProb_y_J2-0.5}}
\end{figure}

The percolation probability in the vertical direction $r_y$, on the other hand, is a much more feature-rich function in $\beta$; see Fig.~\ref{fig:wrapProb_y_J2-0.5}(a). Starting at an initially low value for small $\beta$, at first $r_y$ increases in a similar fashion as $r_x$, however, instead of approaching its zero-temperature-value monotonously as $\beta$ increases, it exhibits a maximum followed by a sharp decline, proceeded by a minimum that becomes deeper with system size, and finally approaches a system-size dependent constant around $0.8$ in the large $\beta$ limit.
This constant converges very slowly to one as $L\to\infty$, see the asymptotics in Eq.~\eqref{eq:zeroTempPercProbY} below.

Besides the maximum and the minimum, the low-$\beta$ point at half the maximum height and the $\beta$ at which $r_y$ is halfway between its maximum and minimum are marked by open circles and squares, respectively.  Figure~\ref{fig:wrapProb_y_J2-0.5}(b) shows these special $\beta$-points as a function of $L$, alongside the $\beta$ at which the horizontal wrapping probability $r_x$ is equal to 0.5, and the $\beta$ of the specific-heat maximum. Most notably, the drop of $r_y$ (indicated by its halfway point) is at the same temperature as the specific-heat peak.

These features are easiest understood by working backwards from the $\beta \rightarrow \infty$ limit, i.e., from zero temperature, to lower $\beta$.  $2^L$ of the $2^L+2^{L/2+1}-n_\text{oc}$ ground states can be described as parallel RWs which wrap in the horizontal direction (see Sec.~\ref{sec:groundStates}). We will disregard the additional $2^{L/2+1}-n_\text{oc}$ ground states, as they are negligible already for the system sizes considered here. Unless the RW ends exactly where it started after $L$ steps, its corresponding spin configuration consists of horizontally wrapping stripes which spiral in the vertical direction. Therefore, in the absence of defects it also percolates in the vertical direction. As discussed in Sec.~\ref{sec:groundStates}, the RW has to consist of an even number of steps in either direction, and therefore reduces the number of allowed RWs to $2^{L-1}$. Further, there are ${L \choose L/2}$ one-dimensional RWs of length $L$ with zero end-to-end distance, and thus the plateau in $r_y$ at zero temperature is
\begin{equation}
r_y (L,T=0) = 1 - \frac{{L \choose L/2}}{2^{L-1}} \stackrel{(\ast)}{=} 1 - \sqrt{\frac{8}{\pi L}} e^{-\frac{1}{4L} + O\left(\frac{1}{L^3}\right)}, \label{eq:zeroTempPercProbY}
\end{equation}
which is in good agreement with the data of Fig.~\ref{fig:wrapProb_y_J2-0.5}(a) for large $\beta$ (dashed lines). In $(\ast)$ the Stirling approximation $n! \sim \sqrt{2\pi n} (n/e)^n \exp[1/(12n)-O(1/n^3)]$ was applied and clearly shows that $r_x$ approaches $1$ with an order of convergence $1/2$ as $L\rightarrow \infty$. Following above RW argument, one would expect that the RW in each horizontal `rotation' spirals on average $\sim\sqrt{L}$ up- or downwards, and that there are in total $\sim\sqrt{L}$ many stripes of length $\sim L^{3/2}$ percolating vertically. On each stripe a single excitation suffices to break its vertical percolation property, and once there is at least one such defect on all $\sim \sqrt{L}$ stripes the configuration looses its percolation property. The number of defects required scales as $~\sqrt{L} \ln{L}$ (coupon collector's problem), which grows slower than the linear system size, and therefore the vertical percolation is broken without restoring the $\eta$-symmetry yielding the minimum in $r_y$ which becomes deeper with system size.

From the high-temperature direction the percolation probability increases with $\beta$. When the $\eta$-symmetry is broken and the vertical direction is pinned, simple (non-spiraling) structures percolating in the vertical direction have to be broken up and have to be replaced by aforementioned spirals. As the spirals are only stable at much lower temperatures, the $\eta$-symmetry-breaking transition is reflected in a sharp decrease in $r_y$. Numerically this becomes visible by the decrease in $r_y$ (halfway points) which as noted above coincide with the peaks in the specific heat, viz.~Fig.~\ref{fig:wrapProb_y_J2-0.5}(b).

\section{Direct comparison of PA simulations of different aspect ratios with literature data}\label{app:directComparisonSimulations}

\begin{figure}[tb!]
  \includegraphics{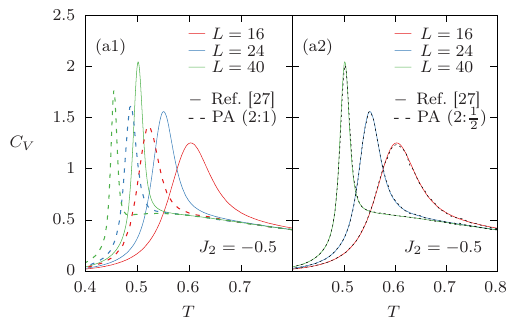}
  \includegraphics{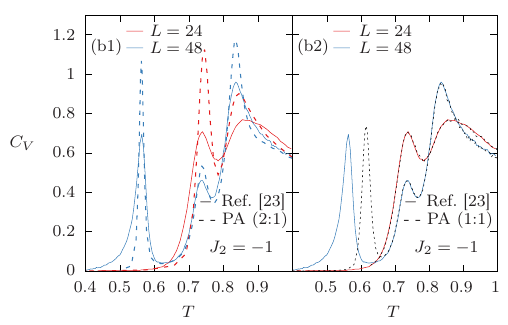}
  \caption{Direct comparison of simulation data from Refs.~\cite{Azhari2025,Zukovic2022} (solid lines) and data from our PA simulations (dashed lines). The ratio in parentheses indicates the used aspect ratio of the lattice. (a1,a2) Comparison to Fig.~5 of Ref.~\cite{Azhari2025} with $J_2=-0.5$. (b1,b2) Comparison to Fig.~6 of Ref.~\cite{Zukovic2022} with $J_2=-1$. In both cases, the PA data in the top row uses the same aspect ratio as in the main text and the PA data in the bottom row the one of the respective reference.\label{fig:literatureComp}}
\end{figure}

In Fig.~\ref{fig:literatureComp} we directly compare our simulation data to those of Refs.~\cite{Azhari2025,Zukovic2022} as further evidence of the quoted aspect ratios in the main text. In both cases, the top row shows data from our PA simulations simulations (where $L_x:L_y = 2\!:\!1$ with $L_y = L$) of the same system size $L$ as quoted in Refs.~\cite{Azhari2025,Zukovic2022}, which differ significantly.  Conversely, the bottom row shows additional PA data using simulations with the aspect ratios $L_x$:$L_y$ being 2:$1/2$ in Fig.~\ref{fig:literatureComp}(a2) and 1:1 in Fig.~\ref{fig:literatureComp}(b2). When comparing the data for $J_2=-0.5$ the agreement with Ref.~\cite{Azhari2025} is nearly perfect with only very small differences visible for the largest system size $L=40$.  For $J_2=-1$, the data for $L=24$ are also in very good agreement with those of Ref.~\cite{Zukovic2022}, cf.\ Fig.~\ref{fig:literatureComp}(b2). For $L=48$, we still find good agreement for the high-temperature region, while the lowest-temperature peak differs significantly. We attribute this difference to the difficulty in simulating and equilibrating the $J_2=-1$ system at these low temperatures that was also visible in our data in the main text (see Fig.~\ref{fig:takeover_-1_L48} and the surrounding discussion). 

\section{Analogous behavior in simple exactly solvable models} \label{app:analogousBehaviorInSimpleModels}
\subsection{Specific heat of the two-dimensional gonihedric model on the square lattice}\label{app:gonihedricModel}

The $d=2$ gonihedric Ising model is given by \cite{savvidy:94}
\begin{equation}
  \mathcal{H} = -\kappa \sum_{\langle ij \rangle} \sigma_i\sigma_j + \frac{\kappa}{2} \sum_{[ ik ]} \sigma_i\sigma_k-\frac{1-\kappa}{2} \sum_{[ijkl]} \sigma_i\sigma_j\sigma_k\sigma_l, \label{eq:gonihedricModel}
\end{equation}
where $\sigma_i$ are Ising spins on a square lattice, and $\sum_{\langle ij \rangle}$, $\sum_{[ ik ]}$, $\sum_{[ijkl]}$ denote the sums over nearest neighbors, next-nearest neighbors, and four-spin plaquettes, respectively.
For $\kappa=1$ this is equivalent to the $J_1$-$J_2$ model with $J_2 = -1/2$, albeit on the square lattice.
For $\kappa=0$ (i.e., the plaquettes-only case) it can be solved analytically for the infinite system and finite systems with free, periodic, and mixed boundary conditions and has no thermodynamic singularities~\cite{Espriu2004,Mueller2017}. Similar to the honeycomb model discussed in the main text, the specific heat shows a sharp peak when using periodic boundary conditions, which grows for small $L$ and then becomes smaller again. In the thermodynamic limit, and when using free boundary conditions, the peak disappears.

\begin{figure}[tb!]
  \includegraphics{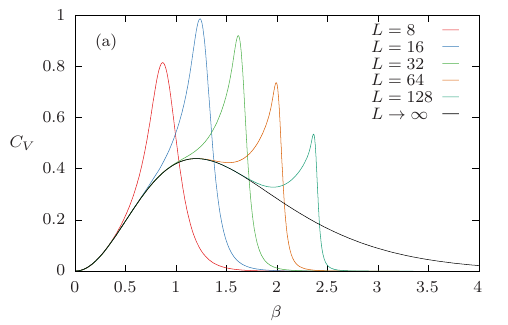}
  \includegraphics{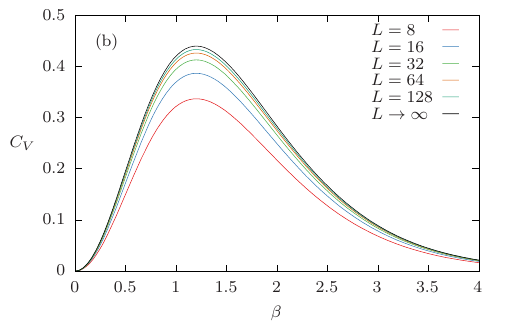}
  \caption{Exact specific heat for the two-dimensional gonihedric plaquette model given by Eq.~(\ref{eq:gonihedricModel}) with $\kappa=0$ for different system sizes, with (a) PBC and (b) FBC. \label{fig:gonihedricCv}}
\end{figure}

The partition function of an $L_x \times L_y$ system with periodic boundary conditions is given by~\cite{Espriu2004,Mueller2017}
\begin{align}
Z ^{(\text{per})}_{L_x,L_y} (\beta) = &\frac 1 2 \left[ 2\cosh(\beta) \right]^{L_x L_y} \times \\ & \sum _{k=0}^{L_x} \binom{L_x}{k} \big(\tanh^{L_x-k}(\beta)+\tanh^k(\beta)\big)^{L_y}, \nonumber
\end{align}
and for free boundary conditions by~\cite{Mueller2017}
\begin{equation}
   Z ^{(\text{free})}_{L_x,L_y} (\beta) = 2^{L_x L_y} \cosh ^{(L_x-1) (L_y-1)}(\beta).
\end{equation}

From this it is straightforward to obtain the specific heat $C_V = \beta^2 L_x^{-1} L_y^{-1} \partial^2 \ln Z / \partial \beta^2$ which in the thermodynamic limit is \begin{equation} C_V (\beta) = \frac {\beta^2} {\cosh^2 \beta}.  \end{equation} Figure~\ref{fig:gonihedricCv} shows the numerically evaluated specific heat for periodic and free boundary conditions for different lattice sizes $L=L_x=L_y$.

\subsection{Binder parameter of the one-dimensional Ising model}\label{app:binderParam1DIsing}

The partition function for the periodic Ising chain of length $L$ as a function of inverse temperature $\beta$ and external field $h$ can be written as (setting $J=1$)
\begin{equation}
    Z_L (\beta,h) = \left[\lambda_{+}(\beta,h)\right]^L + \left[\lambda_{-}(\beta,h)\right]^L,
\end{equation}
where $\lambda_{+}$ and $\lambda_{-}$ are given by
\begin{equation}
    \lambda_{\pm} = e^\beta \left(\cosh(\beta h) \pm \sqrt{\sinh^2(\beta h)+e^{-4\beta}} \right).
\end{equation}

\begin{figure}[ht!]
    \includegraphics{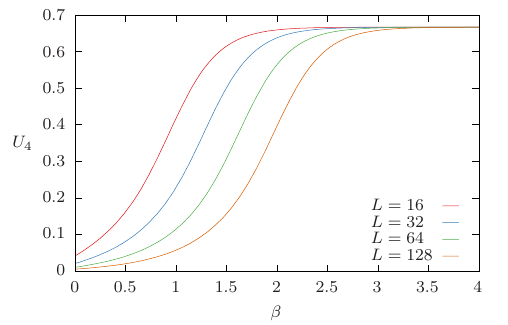}
    \caption{Binder parameter $U_4$ for a periodic Ising chain as a function of inverse temperature $\beta$ for different lengths $L$.\label{fig:binderParam1DIsing}}
\end{figure}

Moments of the magnetization can be calculated by $\langle M^k \rangle = \beta^{-k} [Z_L(\beta,h)]^{-1} \partial^k / \partial h^k Z_L (\beta,h)$, such that at $h=0$ the Binder parameter $U_4 = 1 - \langle M^4 \rangle / 3 \langle M^2 \rangle^2$ can be easily evaluated. Figure~\ref{fig:binderParam1DIsing} shows the numerically evaluated Binder parameter for different chain lengths~$L$ which resembles Fig.~\ref{fig:fss_-0.5}(a).

\vspace{10cm}

\end{document}